\documentclass[twocolumn]{aastex631}

\usepackage[T1]{fontenc}
\usepackage{amsmath}
\usepackage{amssymb}
\usepackage{bm}
\usepackage{booktabs}
\usepackage{graphicx}

\graphicspath{{figures/}}
\newcommand{\Mearth}{M_{\oplus}}
\newcommand{\Mjup}{M_{\rm J}}
\newcommand{\Msun}{M_{\odot}}
\newcommand{\RH}{R_{\rm H}}
\newcommand{\SEMM}{solids-enhanced minimum-mass}

\newcommand{\inputcola}[1]{\parbox[t]{0.22\textwidth}{\raggedright #1}}
\newcommand{\inputcolb}[1]{\parbox[t]{0.40\textwidth}{\raggedright #1}}
\newcommand{\inputcolc}[1]{\parbox[t]{0.33\textwidth}{\raggedright #1}}

\shorttitle{Constraints from PDS 70 c and SR 12 c}
\shortauthors{Mosqueira}

\hypersetup{
  pdftitle={PDS 70 c and SR 12 c: Observational Constraints on Giant-Planet and Satellite Formation},
  pdfauthor={I. Mosqueira}
}

\begin{document}

\title{PDS~70~c and SR~12~c: Observational Constraints on Giant-Planet and Satellite Formation}

\author{I. Mosqueira}
\affiliation{Department of Physics and Astronomy, San Jos\'e State University, 1 Washington Square, San Jos\'e, CA 95192, USA}
\email{Ignacio.Mosqueira@sjsu.edu}

\begin{abstract}
PDS~70~c and SR~12~c are the only bound planetary-mass objects with secure cold submillimeter disk detections.  Together, these systems constrain giant-planet growth and satellite formation.  The PDS~70 planets exhibit remarkable parallels to the Jupiter--Saturn pair in our Solar System.  Both PDS~70 planets accrete within one shared gap, which links their final masses, the material reaching each Hill sphere, and the properties of the circumplanetary disk.  Planetary torques deplete the finite interplanetary reservoir, causing circumplanetary supply to decline as the protoplanets open a circumstellar gap.  SR~12~c separates the planetary-growth and satellite-formation timescales: gas and solids survive even though its current mass-growth timescale is $(1.9\pm0.6)\times10^9$~yr.  For PDS~70~c, the 855-$\mu$m flux implies $0.007$--$0.031\,\Mearth$ of dust at 26~K in the optically thin limit.  A fully dust-dominated, uniform 22--26-K optically thick emitter has an equivalent coplanar radius of $0.58$--$0.66$~au, while a fiducial radial temperature profile yields an equivalent radius of approximately 0.46~au.  The continuum constraints overlap the 0.5--1.5~au circularization range estimated by ballistic calculations of late-stage gap-fed inflow.  We find that the PDS~70 constraints are consistent with our \SEMM{} satellite-formation model (Mosqueira \& Estrada 2003a,b, submitted in 2001).  Thus the observations provide strong support for a quiescent, solids-enhanced satellite-forming environment, coupled in the early stages to planetary-gap evolution.
\end{abstract}


\section{Introduction}\label{sec:intro}

The PDS~70 system connects a directly imaged pair of accreting giant planets to a spatially localized circumplanetary continuum source while the planets remain embedded in their natal disk.  PDS~70~b lies near 22~au inside the large dust-depleted cavity \citep{Keppler2018}; PDS~70~c lies near 34.5~au, and both planets have been detected in H$\alpha$ emission \citep{Haffert2019}.  The stellar mass is approximately $0.76\,\Msun$, the distance is 112.4~pc, and the system age is $5.4\pm1.0$~Myr \citep{Muller2018,Wang2021}.  The system can therefore be viewed as an analogue of the Jupiter--Saturn architecture in the sense most relevant here: two neighboring giant planets and their circumplanetary environments are embedded in one large-scale gap.

The key motivational observation was the 855-$\mu$m Atacama Large Millimeter/submillimeter Array (ALMA) image of \citet{Benisty2021}.  At an angular resolution of about 20~mas, a compact and detached source was found at the predicted position of PDS~70~c.  The inferred source radius is smaller than approximately 1.2~au, and the optically thin dust masses obtained for the adopted grain-size distributions span approximately $0.007$--$0.031\,\Mearth$ at 26~K.  The source lies well inside the planet's Hill sphere and is spatially associated with the planet rather than the circumstellar ring.  Its interpretation as a satellite-forming circumplanetary disk (CPD) is therefore well founded.

Two lines of work anticipate the physical picture advanced here.  \citet{BalbusHawley1998} and \citet{Hawley1999} show that hydrodynamic turbulent fluctuations in the presence of Keplerian shear decay through Coriolis-mediated epicyclic energy exchange coupled to the turbulence cascade (see Appendix~\ref{app:les}).  A second line explains the masses, compositions, and survival of the regular satellites of Jupiter and Saturn with a quiescent, solids-enhanced circumplanetary disk \citep{MosqueiraEstrada2003a,MosqueiraEstrada2003b,Estrada2009,Mosqueira2010SSR}.  PDS~70~c and SR~12~c provide direct examples of quiescent conditions in planet-forming circumplanetary environments, while the settling surveys discussed below establish that weak stirring is common in a broader range of disks.

The vertical distribution of millimeter grains directly constrains the turbulent stirring available to counter dust settling.  In HL~Tau, \citet{Pinte2016} found that the inferred thin dust layer is consistent with $\alpha\sim10^{-4}$.  Direct ALMA imaging of 12 edge-on disks subsequently showed significant vertical settling of large grains \citep{Villenave2020}.  Oph~163131 provides an especially stringent case: its millimeter-grain scale height is about an order of magnitude below that of the micron-sized grains, resulting in $\alpha\lesssim10^{-5}$ \citep{Villenave2022}.  A uniform analysis of 33 disks found vertically settled outer disks in every one of the 23 systems for which the thickness could be constrained \citep{Villenave2025}.  Typical upper limits on the corresponding inferred coefficient $\alpha$ are $\lesssim10^{-3}$; in several systems the inferred stirring is too weak for turbulence to be the principal driver of accretion.

Molecular-line widths independently constrain gas motions in the layers and radial regions sampled by each transition.  \citet{Flaherty2020} obtained upper limits of $0.08\,c_s$ in MWC~480 and $0.12\,c_s$ in V4046~Sgr, where $c_s$ is the local sound speed.  The same analysis inferred nonthermal CO line broadening of $0.25$--$0.33\,c_s$ in DM~Tau.  However, for this detection to translate into observational support for an $\alpha$ disk planet-formation model, the following hurdles would need to be cleared: first, the measured broadening would have to be tied to a correlated radial--azimuthal stress; second, one would have to establish that the system is in the planet-forming phase; and third, the CO-emitting region would need to be shown to overlap with the disk's planet-forming region.  Taken together, these observations make a robust case that weak turbulence should be regarded as the empirical prior in the disk regions for which settling or line-width constraints exist.  (See Appendix~\ref{app:ai-bias} for an example of publication-record confirmation bias inherited by an artificial-intelligence assistant that reversed the evidentiary burden just described.)

Since the original observations in 2021, PDS~70~c has been recovered at multiple frequencies and epochs \citep{Fasano2025,DominguezJamett2025}.  The independent 0.88-mm detection of a disk around SR~12~c shows that the PDS~70~c continuum reservoir is not an isolated planetary-mass observation; the two systems lie on the extrapolated 0.88-mm disk--host relation \citep{Wu2020,Wu2022}.  The low Band~4--Band~7 spectral index has motivated both an optically thick dust-ring interpretation and consideration of a possible variable non-dust contribution \citep{DominguezJamett2025,Casassus2026,Shibaike2026}.  Molecular-line observations have also mapped the circumstellar gas in increasing detail \citep{Law2024,Rampinelli2024}.  These measurements sharpen the radiative-transfer limits but do not yet supply a direct gas mass for the compact CPD.

The central claim here is broad.  On the circumstellar side, the two PDS~70 giants share one gap; their tidal torques deplete the accessible gas, limit their final masses, change the angular momentum reaching their Hill spheres, and ultimately exhaust circumplanetary delivery from the finite interplanetary reservoir in the isolated limit.  On the circumplanetary side, the delivery of circumstellar solids and gas constrains the compositional budget available to build moons.

Section~\ref{sec:obs} establishes the observational constraints on PDS~70~c and SR~12~c, including the optically thin and optically thick continuum limits.  Section~\ref{sec:rc} relates the inflow angular momentum to the CPD radial scale.  Section~\ref{sec:gapdepletion} gives the torque-driven depletion of the finite circumstellar reservoir in the common planetary region.  Section~\ref{sec:ballistic} maps a finite-height, gap-fed angular-momentum distribution into a normalized CPD deposition profile while separating Hill-sphere entry, retention, and mass normalization.  Section~\ref{sec:les} characterizes the transient stream--disk response and its decay (see Appendix~\ref{app:les} for the supporting calculation).  Section~\ref{sec:satellites} develops the implications for the \SEMM{} chronology, regular-satellite formation, and planetesimal-fragment delivery.  Appendix~\ref{app:scalings} collects the adopted quantities and convenient numerical scalings.

\section{Observational Constraints on Circumplanetary Disks}\label{sec:obs}

\subsection{The two-planet architecture and the common gap}

\begin{figure*}[t]
\centering
\includegraphics[width=0.95\textwidth]{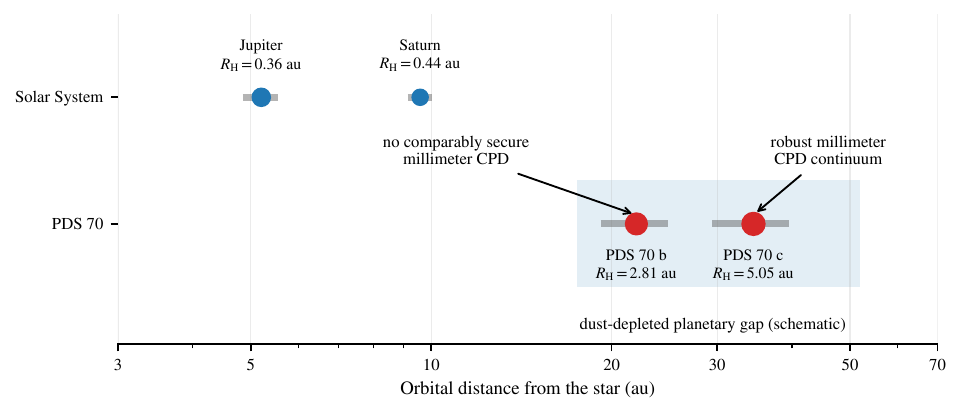}
\caption{Radial architecture of Jupiter--Saturn and PDS~70~b--c.  The plotted PDS~70 Hill radii use the fiducial masses $M_b=5\,\Mjup$ and $M_c=7.5\,\Mjup$.  A compact millimeter continuum source is robustly associated with PDS~70~c.  Recent Band~6 analyses report tentative emission near PDS~70~b, but there is still no comparably secure multi-band CPD detection at b.}
\label{fig:architecture}
\end{figure*}

Figure~\ref{fig:architecture} places the PDS~70 planets alongside Jupiter and Saturn on the same logarithmic axis.  In both systems, two neighboring giant planets control a shared radial region of the parent disk.

New dynamical analyses give upper mass limits of approximately $4.9\,\Mjup$ for b and $13.6\,\Mjup$ for c \citep{Trevascus2025}.  Because the continuum scalings below depend only weakly on the planet mass, we use the $4$--$12\,\Mjup$ range considered by \citet{Shibaike2026} for PDS~70~c.

\subsection{The continuum source at PDS~70~c}

\begin{figure}[t]
\centering
\includegraphics[width=\columnwidth]{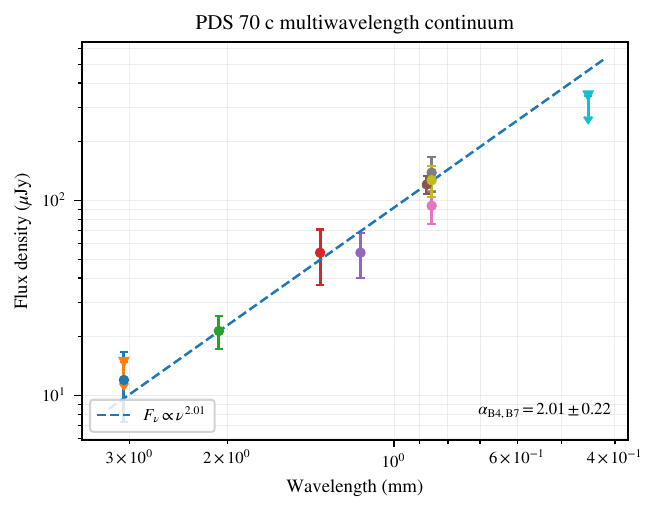}
\caption{Multiwavelength continuum measurements at PDS~70~c.  Detections and upper limits are from Table~\ref{tab:sed}.  The dashed line is the Band~4--Band~7 spectral slope reported by \citet{DominguezJamett2025}, $F_\nu\propto\nu^\alpha$ with $\alpha_{\rm B4,B7}=2.01\pm0.22$.  It uses their 2.07-mm ($21.4\pm4.1\,\mu{\rm Jy}$) and 0.873-mm ($121\pm13\,\mu{\rm Jy}$) measurements, but the quoted index uncertainty is their published result rather than an independent propagation from these rounded flux errors.  The low spectral index motivates an optically thick dust ring; the Band~9 upper limit and reported variability motivate consideration of a mixed or non-dust component.}
\label{fig:sed}
\end{figure}

The original Band~7 detection has been reinforced by multi-epoch measurements at 0.86--1.36~mm and by a Band~4 detection at 2.07~mm \citep{Fasano2025,DominguezJamett2025}.  Table~\ref{tab:sed} lists the measurements assembled for the spectral-energy-distribution calculation by \citet{Shibaike2026}.  The Band~4 and contemporaneous Band~7 measurements give the two-point spectral index
\begin{equation}
F_\nu\propto \nu^\alpha,
\qquad
\alpha_{\rm B4,B7}=2.01\pm0.22,
\label{eq:alpha}
\end{equation}
where $F_\nu$ is the flux density at observing frequency $\nu$ and $\alpha$ is the spectral index.  At 22--26~K, the Band~4--Band~7 interval is not fully in the Rayleigh--Jeans regime.  The appropriate isothermal, optically thick comparison is the finite-temperature Planck slope,
\begin{equation*}
\alpha_B(T)=
\frac{\ln[B_{\nu_7}(T)/B_{\nu_4}(T)]}
{\ln(\nu_7/\nu_4)},
\end{equation*}
which gives $\alpha_B=1.73$ at 22~K and 1.77 at 26~K for the 2.07- and 0.873-mm measurements.  The observed $\alpha=2.01\pm0.22$ remains statistically consistent with these optically thick thermal values.  Finite optical depth, scattering, radial temperature structure, geometry, a mixed emission component, and variability can modify this comparison.  The value and uncertainty are those reported for the Band~4--Band~7 pair in Table~4 of \citet{DominguezJamett2025}.  \citet{Shibaike2026} subsequently quoted $2.01\pm0.19$ for the same pair without documenting a revised uncertainty calculation.  Direct propagation treating the rounded flux errors in Table~\ref{tab:sed} as independent gives $\sigma_\alpha\simeq0.25$; consequently, we retain the primary observational paper's published $0.22$ uncertainty and do not describe it as independently rederived from the tabulated flux errors.  This small uncertainty difference does not change the interpretation of the slope.  The measurements and the reported slope are shown in Figure~\ref{fig:sed}.

\begin{deluxetable*}{cccl}
\tablecaption{Multiwavelength continuum measurements at PDS~70~c.\label{tab:sed}}
\tablehead{
\colhead{ALMA band} & \colhead{$\lambda$ (mm)} & \colhead{$F_\nu$ ($\mu{\rm Jy}$)} & \colhead{Status and source}}
\startdata
3 & 3.07 & $<14.82$ & $3\sigma$ upper limit, original analysis \citep{Doi2024} \\
3 & 3.07 & $12.0\pm4.7$ & marginal reanalysis \citep{DominguezJamett2025} \\
4 & 2.07 & $21.4\pm4.1$ & detection \citep{DominguezJamett2025} \\
6 & 1.36 & $54\pm17$ & multi-epoch measurement \citep{Fasano2025} \\
6 & 1.15 & $54\pm14$ & multi-epoch measurement \citep{Fasano2025} \\
7 & 0.855 & $86\pm16$ & unresolved-source reference flux (peak intensity), discovery analysis \citep{Benisty2021} \\
7 & 0.873 & $121\pm13$ & 2023 measurement \citep{DominguezJamett2025} \\
7 & 0.856 & $94\pm18$ & 2016 and 2019 epochs \citep{Fasano2025} \\
7 & 0.856 & $139\pm28$ & 2021 epochs \citep{Fasano2025} \\
7 & 0.856 & $127\pm23$ & 2023 epochs \citep{Fasano2025} \\
9 & 0.447 & $<345$ & $3\sigma$ upper limit \citep{DominguezJamett2025} \\
\enddata
\tablecomments{Follow-up values follow the compilation in \citet{Shibaike2026}; the discovery reference flux is from \citet{Benisty2021}.  Band~7 epoch-to-epoch differences are not used here to infer a dust-mass evolution timescale.  Time-differential analyses have reported possible short-timescale variability \citep{Casassus2026}, which is relevant to the emission mechanism but not to the existence of a compact source associated with c.}
\end{deluxetable*}

The low spectral index has produced two current interpretations.  \citet{DominguezJamett2025} examined free--free and shock-related emission in the circumplanetary environment, and \citet{Casassus2026} reported possible hourly Band~7 variability.  \citet{Shibaike2026} showed that an optically thick dust ring can fit the multiwavelength continuum and can reach a high local solids column.  The dust-dominated optically thin and optically thick limits are shown separately in Section~\ref{sec:continuum}.

\subsection{What the PDS~70~c continuum requires}\label{sec:continuum}

The continuum can be interpreted without first assuming that it is optically thin.  For an axisymmetric disk inclined by $i$, the flux density is
\begin{equation}
F_\nu = \frac{2\pi\cos i}{d^2}
\int_{r_{\rm in}}^{r_{\rm out}}
B_\nu[T(r)]\left[1-e^{-\tau_\nu(r)/\cos i}\right]r\,dr,
\label{eq:fluxgeneral}
\end{equation}
where $d$ is the source distance, $r$ is the planet-centered cylindrical radius, $r_{\rm in}$ and $r_{\rm out}$ bound the emitting region, and $T(r)$ is the dust temperature profile.  The inclination is $i=0$ for a face-on disk.
In this expression,
\begin{equation}
B_\nu(T)=\frac{2h\nu^3}{c^2}
\left(e^{h\nu/k_{\rm B}T}-1\right)^{-1},
\label{eq:planck}
\end{equation}
is the Planck function, with $h$, $c$, and $k_{\rm B}$ denoting the Planck constant, speed of light, and Boltzmann constant.  The quantity $\tau_\nu=\kappa_\nu\Sigma_{\rm d}$ is the vertical absorption optical depth in the absorption-only limit; $\kappa_\nu$ is the corresponding opacity per unit dust mass and $\Sigma_{\rm d}$ is the dust surface density.  Scattering modifies the emergent intensity and the relation between spectral index and optical depth \citep{Birnstiel2018}.  Thus optically thin emission measures dust mass for an adopted absorption opacity and temperature, while optically thick emission constrains the emitting area conditional on its temperature structure and dust fraction; a column constraint additionally requires a specified opacity and transfer model.

\subsubsection{Optically thin limit: a satellite-scale dust mass}\label{sec:thin}

For $\tau_\nu\ll1$ and a single dust temperature $T_{\rm d}$, Equation~(\ref{eq:fluxgeneral}) becomes
\begin{equation}
M_{\rm d}=\frac{F_\nu d^2}{\kappa_\nu B_\nu(T_{\rm d})}.
\label{eq:thinmass}
\end{equation}
Here $M_{\rm d}$ is the total emitting dust mass and $T_{\rm d}$ is its adopted single temperature.  The opacity depends on the grain-size distribution,
\begin{equation}
\kappa_\nu=
\frac{\displaystyle\int_{a_{\rm min}}^{a_{\rm max}}
\pi a^2 Q_\nu(a)n(a)\,da}
{\displaystyle\int_{a_{\rm min}}^{a_{\rm max}}
(4\pi/3)\rho_\bullet a^3 n(a)\,da}.
\label{eq:kappa}
\end{equation}
Here $a_{\rm min}$ and $a_{\rm max}$ are the minimum and maximum grain radii, $Q_\nu(a)$ is the absorption efficiency of a grain of radius $a$, $n(a)$ is the differential grain-size distribution, and $\rho_\bullet$ is the material density of the grains.
Using the reference flux $F_{855}=86\pm16\,\mu{\rm Jy}$, $d=112.4$~pc, and the $T_{\rm d}=26$~K temperature estimate adopted by \citet{Benisty2021}, we obtain the following nominal-flux masses from the two representative Disk Substructures at High Angular Resolution Project (DSHARP) opacities for 1-$\mu$m and 1-mm grains \citep{Birnstiel2018}:
\begin{align}
\kappa_{855}=0.79~{\rm cm^2~g^{-1}}&:
&M_{\rm d}&=0.031\,\Mearth,\\
\kappa_{855}=3.63~{\rm cm^2~g^{-1}}&:
&M_{\rm d}&=0.0068\,\Mearth\simeq0.007\,\Mearth.
\end{align}
Callisto has a mass of $0.0180\,\Mearth$, and the four Galilean satellites together have a mass of approximately $0.066\,\Mearth$.  The optically thin inferences therefore place the millimeter-emitting dust in the regular-satellite mass range.  At their assumed temperature of 26~K, the multi-epoch analysis of \citet{Fasano2025} gives approximately $0.008$--$0.063\,\Mearth$, again spanning Callisto's mass and approaching the total Galilean mass.

\begin{figure}[t]
\centering
\includegraphics[width=\columnwidth]{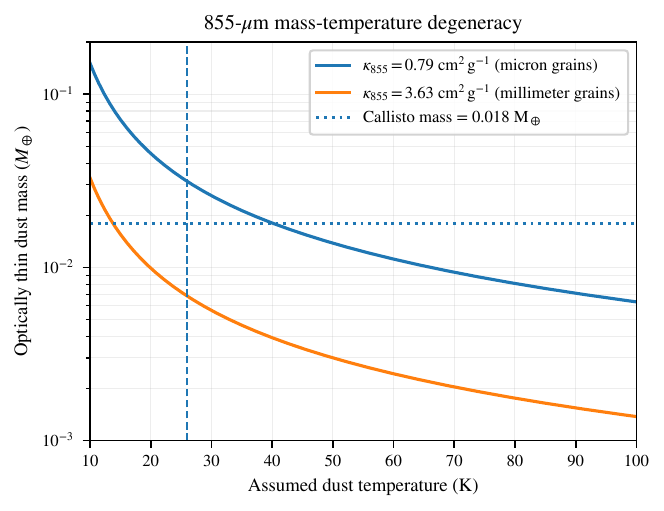}
\caption{Optically thin 855-$\mu$m dust mass as a function of the adopted dust temperature, calculated with the nominal $86\,\mu{\rm Jy}$ flux and 112.4-pc distance used by \citet{Benisty2021}.  The curves use their 1-$\mu$m and 1-mm grain opacities, $\kappa_{855}=0.79$ and $3.63~{\rm cm^2~g^{-1}}$, respectively; their adopted 26-K temperature gives $0.031$ and $0.0068\,\Mearth$, the latter reported as approximately $0.007\,\Mearth$.  The measured flux uncertainty is $\pm16\,\mu{\rm Jy}$ and propagates linearly into these thin-mass values.  The dotted line is Callisto's mass.  The vertical shaded interval shows the $0.008$--$0.063\,\Mearth$ multi-epoch range reported by \citet{Fasano2025} at their assumed temperature of 26~K; it is not temperature independent.}
\label{fig:massT}
\end{figure}

Figure~\ref{fig:massT} shows the temperature dependence of Equation~(\ref{eq:thinmass}).  The inferred mass decreases as the assumed temperature increases.  An imposed heating term would change the mass inferred from a fixed measured flux.  However, our favored quiescent-disk framework does not require a heating term.

If only a fraction $f_{\rm d}(\nu)$ of a measured flux is dust emission, the thin mass is reduced by the same factor,
\begin{equation}
M_{\rm d}=f_{\rm d}(\nu)
\frac{F_\nu d^2}{\kappa_\nu B_\nu(T_{\rm d})}.
\label{eq:mixedmass}
\end{equation}

\subsubsection{Optically thick limit: emitting area and conditional solids column}\label{sec:thick}

If the dust is optically thick, the flux no longer yields a total mass.  For a fully dust-dominated, uniform, isothermal emitting region, saturated intensity provides a conditional minimum projected area,
\begin{equation}
A_{\rm proj}\geq\frac{F_\nu d^2}{B_\nu(T)}.
\label{eq:aproj}
\end{equation}
Defining an equivalent circular projected radius by $A_{\rm proj}=\pi R_{\rm proj}^2$ gives
\begin{equation}
R_{\rm proj}\geq
\left[\frac{F_\nu d^2}{\pi B_\nu(T)}\right]^{1/2}.
\label{eq:rproj}
\end{equation}
For the reference 855-$\mu$m flux and distance above, we evaluate two literature temperature choices.  \citet{Benisty2021} estimated $T=26$~K at 1~au by combining stellar irradiation, planetary irradiation, and viscous heating.  The optically thick ring calculation of \citet{Shibaike2026} instead adopts a 22-K local circumstellar-disk background temperature.  The corresponding nominal projected bounds are
\begin{equation}
R_{\rm proj}\geq
\begin{cases}
0.52~{\rm au}, & T=22~{\rm K},\\
0.46~{\rm au}, & T=26~{\rm K}.
\end{cases}
\label{eq:rprojnumbers}
\end{equation}
If the emitting structure is circular and coplanar with the circumstellar disk, its physical radius is
\begin{equation}
R_{\rm phys}=\frac{R_{\rm proj}}{\sqrt{\cos i}},
\label{eq:rphys}
\end{equation}
which gives $R_{\rm phys}\geq0.66$ and 0.58~au at the circumstellar-disk inclination $i=51.7^\circ$ measured by \citet{Keppler2019}, for 22 and 26~K, respectively.  These are equivalent radii for a fully dust-dominated, uniform, isothermal circular emitter coplanar with the circumstellar disk.  The 26-K value reproduces the 0.58-au result reported by \citet{Benisty2021}; the 22-K value is the corresponding extension using the temperature adopted by \citet{Shibaike2026}.  At fixed temperature and geometry, if dust supplies a fraction $f_{\rm d}(\nu)$ of the observed flux, the isothermal radius scales as
\begin{equation}
R_{\rm phys}[f_{\rm d}(\nu)]
=f_{\rm d}(\nu)^{1/2}R_{\rm phys}[f_{\rm d}(\nu)=1].
\label{eq:thickdustfraction}
\end{equation}
Thus $f_{\rm d}=0.5$ gives approximately 0.46 and 0.41~au at 22 and 26~K, respectively.

For a saturated non-isothermal disk, Equation~(\ref{eq:fluxgeneral}) instead becomes
\begin{equation}
f_{\rm d}(\nu)F_{\nu,{\rm obs}}
=\frac{2\pi\cos i}{d^2}
\int_{r_{\rm in}}^{r_{\rm out}}B_\nu[T(r)]r\,dr.
\label{eq:thickradial}
\end{equation}
As one representative case, the Benisty-normalized passive profile plotted in Figure~\ref{fig:cpdtemperaturemodels}, with the negligible nominal 3-K viscous term omitted, is
\begin{equation}
T^4(r)=(24~{\rm K})^4
+\left[18~{\rm K}\left(\frac{r}{\rm au}\right)^{-1/2}\right]^4.
\label{eq:benistyprofile}
\end{equation}
For $f_{\rm d}=1$, negligible $r_{\rm in}$, the nominal $86\,\mu{\rm Jy}$ flux, and the same distance and inclination, numerical integration gives $r_{\rm out}\simeq0.46$~au.  The result increases to approximately 0.49 and 0.54~au for inner radii of 0.1 and 0.2~au, respectively.  A finite optical depth would increase the required area, whereas a smaller dust fraction would reduce it.  The $\pm16\,\mu{\rm Jy}$ flux uncertainty changes either uniform-temperature radius by approximately $\pm9\%$.  Consequently, the continuum provides conditional equivalent radii rather than a temperature- and emission-independent lower bound on the radius.  The observational upper limit remains $R_{\rm d}\lesssim1.2$~au \citep{Benisty2021}.

In the absorption-only geometry of Equation~(\ref{eq:fluxgeneral}), the optical depth along the line of sight is $\tau_{\nu,{\rm los}}^{\rm abs}=\tau_{\nu,{\rm vert}}^{\rm abs}/\cos i$.  Its unit-optical-depth condition is therefore
\begin{equation}
\tau_{\nu,{\rm los}}^{\rm abs}
=\frac{\tau_{\nu,{\rm vert}}^{\rm abs}}{\cos i}
=\frac{\kappa_\nu\Sigma_{\rm d}}{\cos i}\gtrsim1
\quad\Rightarrow\quad
\Sigma_{\rm d}\gtrsim\frac{\cos i}{\kappa_\nu}.
\label{eq:sigmamin}
\end{equation}
At $i=51.7^\circ$, $\cos i=0.620$, so the two discovery-paper opacities give absorption-only line-of-sight thresholds of approximately $0.17$--$0.78~{\rm g~cm^{-2}}$.  Requiring the vertical absorption optical depth itself to exceed unity would instead give $0.3$--$1.3~{\rm g~cm^{-2}}$; that is a stronger vertical convention, not the least column density implied by line-of-sight thickness.  If scattering contributes appreciably, the emergent intensity depends on both absorption and scattering optical depths, and neither absorption-only threshold is a general observational lower bound.  The optically thick ring model of \citet{Shibaike2026} supplies its own transfer calculation and high solids column, but the total hidden mass remains dependent on the unresolved optical depth and geometry.

The 26-K CPD temperature adopted by \citet{Benisty2021} combines stellar irradiation, planetary irradiation, and a nominal viscous contribution according to
\begin{equation}
T_{\rm CPD}^4=T_{s,{\rm irr}}^4+T_{p,{\rm irr}}^4+T_{\rm vis}^4,
\label{eq:cpdtemperature}
\end{equation}
with $T_{s,{\rm irr}}=24$~K, $T_{p,{\rm irr}}=18$~K, and $T_{\rm vis}=3$~K.  Thus the adopted 26-K scale is essentially an irradiation temperature.

\subsubsection{The passive circumstellar temperature profile}\label{sec:passivetemperature}

\begin{figure*}[t]
\centering
\includegraphics[width=0.98\textwidth]{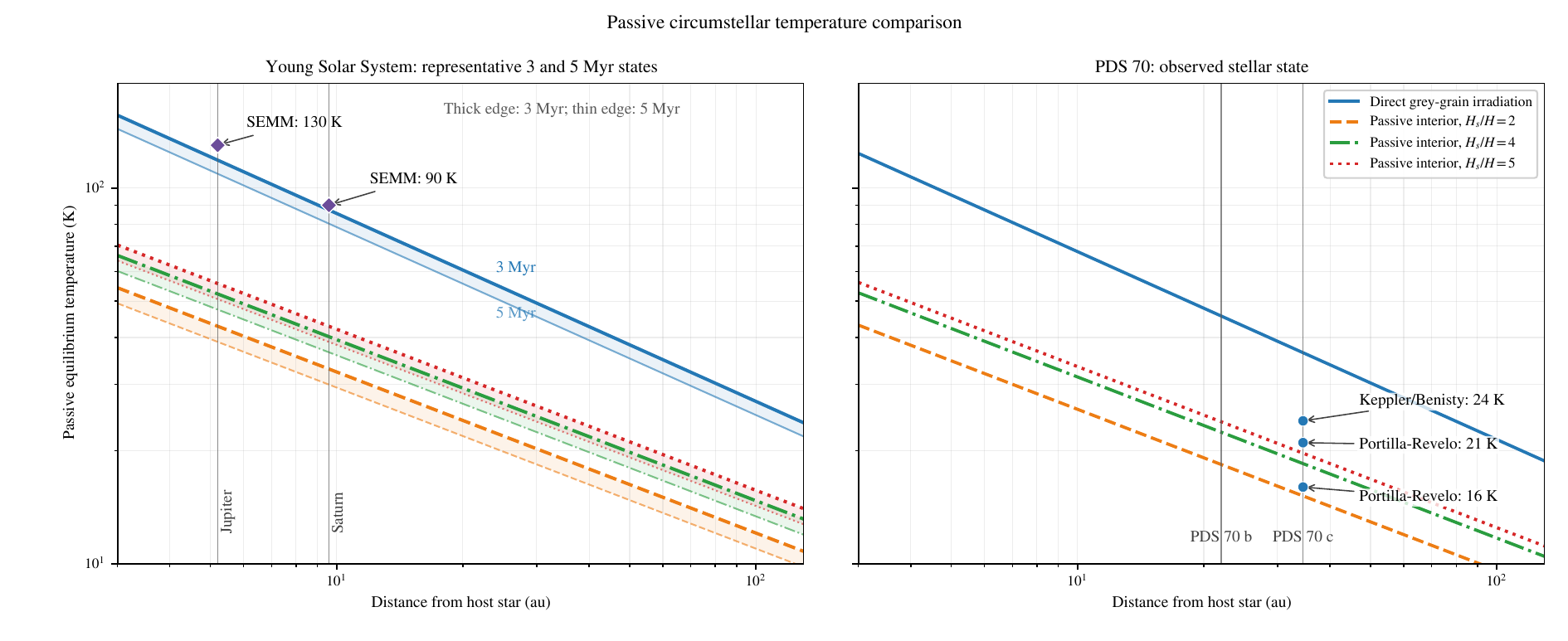}
\caption{Representative passive circumstellar temperature profiles for the young Solar System (left) and PDS~70 (right).  The solid blue curve is the directly illuminated gray-grain equilibrium from Equation~(\ref{eq:directtemperature}); the other curves are passive optically thick interiors with irradiation surfaces at $H_s/H=2$, 4, and 5.  Thick and thin curve edges in the Solar System panel denote representative 3- and 5-Myr stellar states, respectively.  The 130- and 90-K \SEMM{} points were chosen to match the surrounding circumstellar-nebula temperatures at Jupiter and Saturn, so their agreement with a representative circumstellar profile is expected by construction rather than an independent prediction.  The PDS~70~c points show the 16- and 21-K circumstellar midplane models of \citet{PortillaRevelo2023} and the 24-K stellar-irradiation contribution to the CPD temperature adopted by \citet{Benisty2021}.  The curves provide controlled comparisons, not fits to these points.}
\label{fig:passivetemperature}
\end{figure*}

The temperature profile of the circumstellar disk is model dependent.  Figure~\ref{fig:passivetemperature} compares representative passive circumstellar temperature profiles for the young Solar System and PDS~70.  For directly illuminated gray grains, we have
\begin{equation}
T_{\rm dir}(r)=
\left(\frac{L_*}{16\pi\sigma_{\rm SB}r^2}\right)^{1/4},
\label{eq:directtemperature}
\end{equation}
where $L_*$ is the stellar luminosity, $\sigma_{\rm SB}$ is the Stefan--Boltzmann constant, and $r$ is the circumstellar distance.  The optically thick curves use the passive two-layer irradiation geometry of \citet{ChoksiChiangFungZhu2023}.  Here $H$ is the local circumstellar gas pressure scale height, and the irradiation-surface height is $H_s=\chi H$, with $\chi=2$, 4, and 5.  The calculation includes no turbulent, viscous, or inflow heating.  For this geometry, the asymptotic passive-interior profile is
\begin{equation}
T_{\rm int}(r)=
\left[
\frac{\chi}{14}
\left(\frac{k_{\rm B}}{\mu m_{\rm H}GM_*}\right)^{1/2}
R_*^2T_*^4
\right]^{2/7}r^{-3/7},
\label{eq:passiveinteriortemperature}
\end{equation}
where $R_*$ and $T_*$ are the stellar radius and effective temperature, $\mu=2.34$ is the mean molecular weight, and $H_s=\chi H$.  The representative 3-Myr Solar inputs are $(M_*,R_*,T_*,L_*)=(1\,\Msun,1.680\,R_\odot,4330~{\rm K},0.892\,L_\odot)$, and the 5-Myr inputs are $(1\,\Msun,1.434\,R_\odot,4310~{\rm K},0.638\,L_\odot)$.  For PDS~70 we use $(0.76\,\Msun,1.249\,R_\odot,3972~{\rm K},0.35\,L_\odot)$.  The same constructions are applied in both panels.  They are representative rather than well-constrained thermal profiles.  Nevertheless, the curves illustrate the expected circumstellar temperature range at the PDS~70~c orbit and the corresponding Solar System locations of Jupiter and Saturn.

At the orbit of PDS~70~c, Equation~(\ref{eq:directtemperature}) gives approximately 37~K, while the representative passive interiors span approximately 15--20~K.  The 16- and 21-K circumstellar values lie within this passive range, while the 24-K stellar-irradiation contribution used in the CPD estimate of \citet{Benisty2021} lies nearby.  Other published PDS~70 calculations adopt different radial temperature profiles \citep{Muley2019,Bae2019}.

\subsubsection{The passive circumplanetary temperature profile}\label{sec:cpdoutertemperature}
\begin{figure*}[t]
\centering
\includegraphics[width=0.82\textwidth]{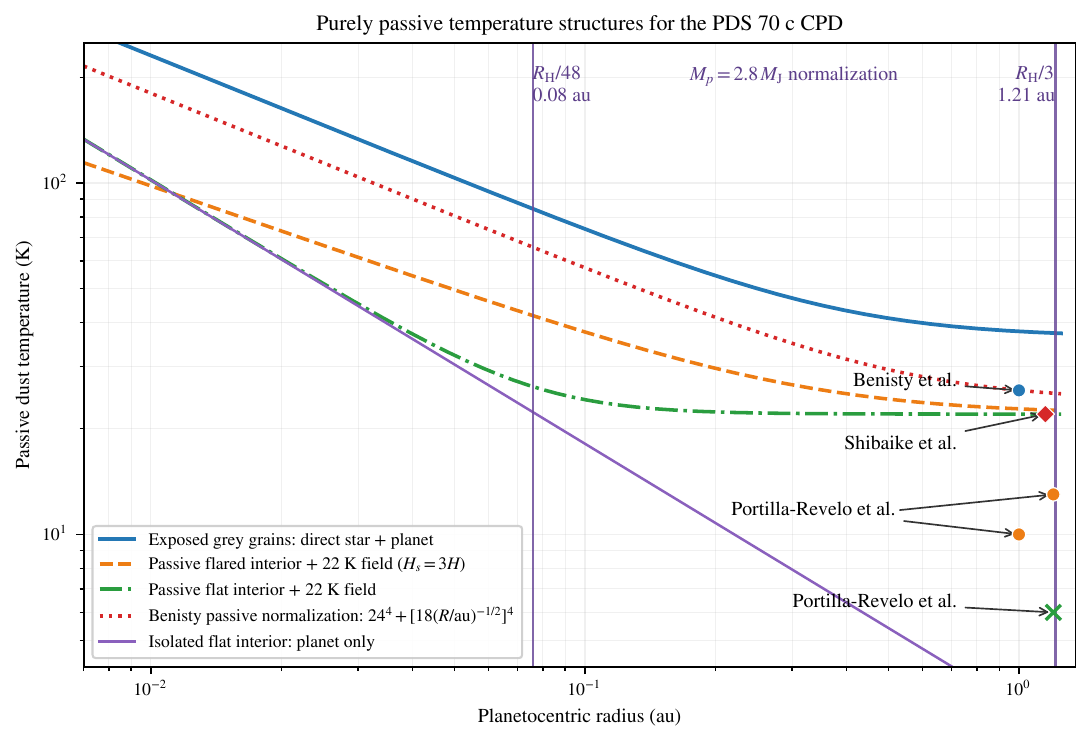}
\caption{Representative purely passive temperature structures for the PDS~70~c circumplanetary disk.  The curves compare directly exposed grains, passive flared and flat optically thick interiors supplied with a 22-K circumstellar background, a profile normalized to the irradiation temperature of \citet{Benisty2021}, and an isolated planet-heated limit.  The author-labeled points show the 26-K estimate of \citet{Benisty2021}, the 22-K external field adopted by \citet{Shibaike2026}, and the 10- and 13-K embedded-disk and 6-K isolated-disk temperatures of \citet{PortillaRevelo2022}.  The solid vertical lines mark $\RH/48$ and $\RH/3$ for the $M_p=2.8\,\Mjup$ normalization used in the plotted thermal calculation.}
\label{fig:cpdtemperaturemodels}
\end{figure*}

A similar passive temperature structure applies to the circumplanetary disk as well, except that the irradiation comes from both the planet and the star (see Figure~\ref{fig:cpdtemperaturemodels}).  Planetary irradiation can maintain a warm inner region, but its contribution decreases with planetocentric distance; in a passive outer disk, the temperature approaches the corresponding circumstellar-disk value at the planet's orbit, $T_{\rm CPD}(r)\rightarrow T_{\rm CSD}(a_p)$, where $T_{\rm CPD}$ and $T_{\rm CSD}$ denote the circumplanetary- and circumstellar-disk temperatures.  This is the construction used in the \SEMM{} calculations, in which the outer circumjovian and circum-Saturnian temperatures were set by the solar-nebula temperatures at Jupiter and Saturn \citep{MosqueiraEstrada2003a,MosqueiraEstrada2003b}.

The directly exposed curve in Figure~\ref{fig:cpdtemperaturemodels} adds the stellar and planetary gray-grain irradiation fields,
\begin{equation}
\begin{aligned}
T_{*,{\rm dir}}&=
\left(\frac{L_*}{16\pi\sigma_{\rm SB}a_p^2}\right)^{1/4},\\
T_{p,{\rm dir}}(r)&=
\left(\frac{L_p}{16\pi\sigma_{\rm SB}r^2}\right)^{1/4},\\
T_{\rm exposed}^4(r)&=T_{*,{\rm dir}}^4+T_{p,{\rm dir}}^4,\\
L_p&=4\pi R_p^2\sigma_{\rm SB}T_p^4.
\end{aligned}
\label{eq:cpdexposedtemperature}
\end{equation}
The optically thick planet-heated interiors use
\begin{equation}
\begin{aligned}
T_{p,{\rm flared}}(r)&=
\left[
\frac{\chi}{14}
\left(\frac{k_{\rm B}}{\mu m_{\rm H}GM_p}\right)^{1/2}
R_p^2T_p^4
\right]^{2/7}r^{-3/7},\\
T_{p,{\rm flat}}(r)&=
0.1^{1/4}T_p\left(\frac{R_p}{r}\right)^{3/4},\\
T_{\rm flared}^4(r)&=(22~{\rm K})^4+T_{p,{\rm flared}}^4(r),\\
T_{\rm flat}^4(r)&=(22~{\rm K})^4+T_{p,{\rm flat}}^4(r).
\end{aligned}
\label{eq:cpdpassivetemperatures}
\end{equation}
Here the flared model uses $\chi=3$ and $\mu=2.34$; the isolated flat-interior curve is $T_{p,{\rm flat}}$ without the 22-K external field.  The adopted inputs are $L_*=0.35\,L_\odot$, $a_p=34.5$~au, $M_p=2.8\,\Mjup$, $R_p=2\,R_{\rm J}$, and $T_p=1054$~K, giving $L_p=4.70\times10^{-5}\,L_\odot$.  These values define representative profiles rather than a fit to a single CPD thermal structure; the Benisty-normalized comparison is given separately by Equation~(\ref{eq:benistyprofile}).

Temperatures of order 250~K apply close to the planet in the optically thick inner part of the \SEMM{} model, whereas the outer satellite-forming disk approaches the temperature of the surrounding circumstellar nebula \citep{MosqueiraEstrada2003a,MosqueiraEstrada2003b,Mosqueira2010SSR}.  A compact warm inner disk and a cold outer disk on the sub-au to au scale allowed by the continuum constraints are consistent with an extended circumplanetary nebula.  The outer-disk temperatures in the \SEMM{} model were selected to match the ambient circumstellar nebula at the locations of Jupiter and Saturn.  Applying the same boundary condition to PDS~70 gives a colder outer CPD because c lies much farther from a cooler star.

\subsection{The present observational status of the PDS~70 circumplanetary disks}

Multi-epoch Band~6 images contain tentative, approximately $3\sigma$ residuals near the expected location of b, with quoted flux densities of order $46$--$59\,\mu{\rm Jy}$ \citep{Fasano2025}.  These residuals do not yet constitute a repeated, multi-band detection like that obtained for c.  Therefore, PDS~70~c has a secure circumplanetary continuum source, while b has no comparably secure millimeter CPD detection.

At the time of the discovery paper, the hoped-for $^{12}$CO and HCO$^+$ gas constraints had not yet been published.  Those observations and related follow-up have now been analyzed, but they do not constrain the gas inventory of the CPD itself.  The CO-isotopologue and HCO$^+$ emitting surfaces mapped by \citet{Law2024} were imaged at an angular resolution of order $0.1''$, or approximately 10~au at PDS~70.  The thermo-chemical reconstruction by \citet{PortillaRevelo2023} measures the circumstellar gas depletion in the planetary gap, and the molecular survey of \citet{Rampinelli2024} maps ring-shaped emission and the chemistry of the parent disk.  These are important measurements of the feeding environment, but the securely detected continuum source at c is smaller than approximately 1.2~au.  The published line data therefore do not spatially isolate gas gravitationally bound to PDS~70~c and do not yield a CPD gas mass or radial gas surface-density profile.

Accretion diagnostics provide a different and narrower statement.  H$\alpha$ emission shows that gas is reaching PDS~70~c, and the possible shock or free--free contribution to the radio continuum would likewise require gas in the stream--CPD interaction region \citep{Haffert2019,DominguezJamett2025,Casassus2026}.  Neither measurement determines how much gas is stored in the rotationally supported CPD.  At present, the observational status is therefore
\begin{equation}
\begin{gathered}
M_{\rm g,c},\quad \Sigma_{\rm g,c}(R),\\
\left(M_{\rm g}/M_{\rm d}\right)_{\!c}
\quad\text{are not directly measured}.
\end{gathered}
\label{eq:pds70_gas_unmeasured}
\end{equation}
Here $M_{\rm g,c}$ is the total gas mass stored in the CPD around PDS~70~c, $\Sigma_{\rm g,c}(R)$ is its gas surface density at planet-centered cylindrical radius $R$, and $(M_{\rm g}/M_{\rm d})_c$ is its total gas-to-dust mass ratio.  The quantity $M_{\rm d,c}$ used below denotes the continuum-inferred dust mass; the subscript c denotes the disk associated with PDS~70~c.  Recent continuum models adopt a gas accretion rate, a gas disk structure, and a dust-to-gas ratio in order to interpret the emission \citep{ShibaikeMordasini2024,Shibaike2026}.

\subsection{The 0.88-mm disk--host relation at planetary masses}\label{sec:hostscaling}

PDS~70~c is not the only bound planetary-mass object with a detected cold submillimeter disk.  \citet{Wu2022} detected unresolved 0.88-mm continuum emission with a flux density of $127\pm14\,\mu{\rm Jy}$ from SR~12~c.  The discovery paper used the Gaia Data Release 2 (DR2) distance of 112.5~pc and described the object as an approximately $11\,\Mjup$ companion at a projected separation of about 980~au.  Because the close host binary has no reliable Gaia DR3 single-star astrometric solution, \citet{Finley2026} instead adopt the mean L1688 cluster distance, $139\pm5$~pc, together with a revised companion mass of $16\pm2\,\Mjup$ and a projected separation of approximately 1200~au.  The dynamical setting is not the same as that of PDS~70: SR~12~c is a wide companion and does not test the common-gap framework emphasized in this paper.  It does, however, provide an independent measurement of the amount of millimeter-emitting material associated with a bound planetary-mass object.

\begin{figure*}[t]
\centering
\includegraphics[width=0.86\textwidth]{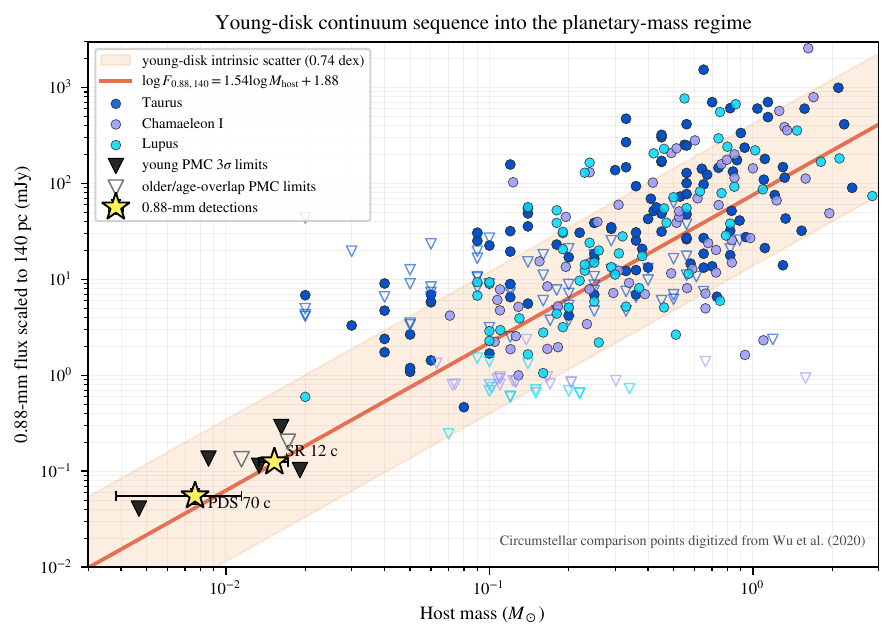}
\caption{The 0.88-mm disk--host relation extended into the planetary-mass regime.  The solid line and shaded region are the 0.88-mm disk--host relation and its 0.74-dex intrinsic scatter from \citet{Wu2020}, with all fluxes scaled to 140~pc using Equation~(\ref{eq:flux140}).  Colored circles and open downward triangles show the Taurus, Chamaeleon~I, and Lupus detections and upper limits adapted from \citet{Wu2020}; filled black and open gray triangles show the young and the older (age-overlap) planetary-mass-companion limits, respectively.  Stars mark PDS~70~c and SR~12~c; their horizontal bars show the adopted $4$--$12\,\Mjup$ and $16\pm2\,\Mjup$ mass ranges, respectively.  The WISPIT~2~b $3\sigma$ point-source limit is $0.041$~mJy when scaled to 140~pc \citep{Facchini2026}.  For PDS~70~c, the 855-$\mu$m discovery flux is used as the 0.88-mm point without a color correction; it scales to $0.055\pm0.010$~mJy.  Using the current $139\pm5$~pc cluster distance and $16\,\Mjup$ mass adopted by \citet{Finley2026}, SR~12~c scales to $0.125\pm0.017$~mJy, compared with approximately $0.12$~mJy predicted by the mean relation at that mass.}
\label{fig:host-scaling}
\end{figure*}

The relevant comparison can be made without converting either continuum flux into a dust mass.  \citet{Wu2020} scaled young-disk 0.88-mm fluxes to a common distance of 140~pc,
\begin{equation}
F_{0.88,140}=F_{0.88,{\rm obs}}
\left(\frac{d}{140~{\rm pc}}\right)^2,
\label{eq:flux140}
\end{equation}
where $F_{0.88,{\rm obs}}$ is the observed 0.88-mm flux density at source distance $d$, and $F_{0.88,140}$ is the flux density scaled to 140~pc.  They fitted the young Taurus, Chamaeleon~I, and Lupus samples with
\begin{equation}
\log_{10}\left(\frac{F_{0.88,140}}{{\rm mJy}}\right)
=1.54\log_{10}\left(\frac{M_{\rm host}}{\Msun}\right)+1.88,
\label{eq:wu_relation}
\end{equation}
where $M_{\rm host}$ is the mass of the central object.  The relation has an intrinsic scatter of 0.74~dex, and the fitted slope and intercept have quoted uncertainties of approximately 0.21 and 0.14, respectively.  At $M_{\rm host}=0.01\,\Msun$, Equation~(\ref{eq:wu_relation}) gives
\begin{equation}
F_{0.88,140}\simeq0.06~{\rm mJy}.
\end{equation}
For PDS~70~c, the discovery flux and distance give
\begin{align}
F_{0.88,140}({\rm PDS~70~c})
&=0.086\left(\frac{112.4}{140}\right)^2\nonumber\\
&=(0.055\pm0.010)~{\rm mJy},
\end{align}
whereas the current cluster-distance estimate for SR~12~c gives
\begin{align}
F_{0.88,140}({\rm SR~12~c})
&=0.127\left(\frac{139}{140}\right)^2\nonumber\\
&=(0.125\pm0.017)~{\rm mJy}.
\end{align}
At the revised mass $16\,\Mjup\simeq0.015\,\Msun$, Equation~(\ref{eq:wu_relation}) predicts
\begin{equation}
F_{0.88,140}^{\rm rel}(16\,\Mjup)\simeq0.12~{\rm mJy}.
\end{equation}
Here the superscript ``rel'' identifies the value predicted by the fitted relation.
The updated SR~12~c point therefore lies close to the extrapolated mean relation.  The quoted uncertainties above reflect propagation of the flux uncertainty for PDS~70~c and of both the flux and cluster-distance uncertainties for SR~12~c; they do not include the 0.74-dex intrinsic scatter of the relation.  The original \citet{Wu2022} parameter set gave $F_{0.88,140}\simeq0.08$~mJy at $11\,\Mjup$, for which the relation predicts approximately $0.07$~mJy; the empirical conclusion is unchanged by the distance and mass revision.  PDS~70~c also lies close to the mean, and both detections are well inside the measured intrinsic scatter.  For WISPIT~2~b, the $45\,\mu{\rm Jy}$ point-source limit at 0.88~mm scales from 133.35 to 140~pc as $F_{0.88,140}<0.041$~mJy; at the estimated mass of $4.9\,\Mjup$, Equation~(\ref{eq:wu_relation}) predicts approximately 0.020~mJy, so the nondetection is consistent with the extrapolated relation \citep{VanCapelleveen2025,Facchini2026}.  Figure~\ref{fig:host-scaling} shows the relation, the six planetary-mass-companion upper limits from the original ALMA survey, the WISPIT~2~b limit, and the two detections.

The published optically thin conversions also give solids inventories of the same order.  For PDS~70~c, the adopted 26-K calculation gives $0.007$--$0.031\,\Mearth$.  For SR~12~c, \citet{Wu2022} obtained approximately $0.012$ and $0.054\,\Mearth$ under their two opacity assumptions at the then-adopted distance of 112.5~pc.  At 139~pc, the SR~12~c values increase by a factor of approximately 1.5, to about $0.02\,\Mearth$ and $0.08\,\Mearth$.  The two detected planetary-mass disks therefore roughly follow the same 0.88-mm disk--host relation defined by disks around higher-mass young hosts.  These detections do not provide an observational basis for treating either planetary or circumplanetary disks as intrinsically mass-depleted samples of astronomical objects.

\subsection{Planetary assembly and circumplanetary-disk survival operate on distinct timescales}\label{sec:sr12growth}

The current accretion state of SR~12~c turns the continuum detection into an evolutionary measurement.  Two independent classes of observations establish the two surviving disk components:
\begin{equation}
\begin{gathered}
\text{submillimeter and mid-infrared continuum}\\
\Longrightarrow \text{surviving solids}.
\end{gathered}
\label{eq:sr12solids}
\end{equation}
\begin{equation}
\begin{gathered}
\text{Balmer continuum, H}\alpha\text{,}\\
\text{and hydrogen-line accretion}\\
\Longrightarrow \text{surviving gas}.
\end{gathered}
\label{eq:sr12gas}
\end{equation}
The ALMA detection and the infrared excess establish a circumplanetary solids reservoir, while the ultraviolet continuum excess, Balmer jump, and hydrogen emission establish that gas is still flowing from the circumplanetary environment onto the companion \citep{Wu2022,Finley2026}.  SR~12~c therefore hosts a circumplanetary disk containing both gas and solids, not merely a residual dust ring.

Ultraviolet imaging with the Hubble Space Telescope (HST) gives
\begin{equation}
\dot M_p=(8\pm2)\times10^{-12}\,\Msun\,{\rm yr^{-1}},
\label{eq:sr12mdot}
\end{equation}
for a companion mass of $M_p=16\pm2\,\Mjup$ \citep{Finley2026}.  This accretion rate was derived from $\dot M_p\simeq1.25R_pL_{\rm acc}/(GM_p)$ using $R_p=1.9\pm0.3\,R_{\rm J}$ and $L_{\rm acc}=(1.65\pm0.19)\times10^{-5}\,L_{\odot}$ \citep{Finley2026}; hence, the quoted uncertainties in $M_p$ and $\dot M_p$ are not independent.  Using the central values and $1\,\Mjup=9.546\times10^{-4}\,\Msun$, we obtain the instantaneous mass-growth time
\begin{align}
t_{\rm grow}
&=\frac{M_p}{\dot M_p}\nonumber\\
&=\frac{16(9.546\times10^{-4})\,\Msun}
{8\times10^{-12}\,\Msun\,{\rm yr^{-1}}}\nonumber\\
&=(1.9\pm0.6)\times10^{9}\,{\rm yr}.
\label{eq:sr12tgrow}
\end{align}
Over another million years at the present rate, the added mass would be only
\begin{align}
\Delta M_p(1\,{\rm Myr})
&=\dot M_p(10^6\,{\rm yr})\nonumber\\
&=(8\pm2)\times10^{-6}\,\Msun\nonumber\\
&=(8.4\pm2.1)\times10^{-3}\,\Mjup,
\label{eq:sr12deltam}
\end{align}
and therefore
\begin{equation}
\frac{\Delta M_p}{M_p}
=(5.2\pm1.7)\times10^{-4}
=(0.052\pm0.017)\%.
\label{eq:sr12fraction}
\end{equation}
The uncertainties in $t_{\rm grow}$ and $\Delta M_p/M_p$ are propagated from the quoted errors in $M_p$, $R_p$, and $L_{\rm acc}$, retaining the dependence $t_{\rm grow}\propto M_p^2/(R_pL_{\rm acc})$; they do not include systematic uncertainty in the adopted magnetospheric-truncation factor.  The uncertainty in $\Delta M_p$ follows directly from the reported $\dot M_p$.  The resulting fraction is consistent with the approximately 0.06\% value quoted by \citet{Finley2026}.  Therefore, growth of the planetary companion is effectively complete even though both gas and solids remain in the circumplanetary environment.

\subsection{The present detection census by tracer}

The present detection census must be separated by tracer.  PDS~70~c and SR~12~c remain the two secure cold 0.88-mm continuum detections around bound planetary-mass companions.  The WISPIT~2~b nondetection adds a $45\,\mu{\rm Jy}$ point-source upper limit at the same wavelength \citep{Facchini2026}.  Two embedded systems provide gaseous or chemical CPD candidates rather than equivalent continuum detections: localized $^{13}$CO emission in AS~209 \citep{Bae2022} and localized HCN and C$_2$H emission in HD~163296 \citep{Izquierdo2026}.  In both cases the central planet is inferred from disk structure or kinematics rather than directly detected.  The James Webb Space Telescope has opened a complementary warm-disk census, including mid-infrared disk signatures around GQ~Lup~B \citep{Cugno2024}, silicate emission around YSES-1~b \citep{Hoch2025}, molecular gas and dust emission around CT~Cha~b \citep{CugnoGrant2025} and Delorme~1~AB~b \citep{Malin2025}, and an infrared excess consistent with a disk around TWA~27~b \citep{Patapis2025}.  Those measurements trace warm micron-sized dust, infrared excess, or molecular gas; they are complementary evidence for circumplanetary material but are not additional points on the cold 0.88-mm disk--host relation.

\subsection{The SR~12~c gas-inventory estimates}

\citet{Wu2022} used the measured accretion rate, $10^{-11.08\pm0.40}\,\Msun\,{\rm yr^{-1}}$, and a remaining disk lifetime of 3~Myr to infer an order-of-magnitude gas inventory
\begin{equation}
M_{\rm g,acc}\sim0.03\,\Mjup.
\label{eq:sr12gasacc}
\end{equation}
Combining this gas inventory with their surface-density-model dust masses, $M_{\rm d}=0.007\,\Mearth$ for millimeter grains and $0.03\,\Mearth$ for micron grains, gave gas-to-dust ratios of 470--3000 and 110--700, respectively.  These are the surface-density-model values reported by \citet{Wu2022}; they differ from the simple matched optically thin estimates above because those estimates use a single mean temperature and opacity conversion.  The revised HST accretion rate is similar to the rate used in that estimate.

\section{Inflow Angular Momentum and the Circumplanetary Scale}\label{sec:rc}

The Hill radius is
\begin{equation}
\RH=a_p\left(\frac{M_p}{3M_*}\right)^{1/3}.
\label{eq:hill}
\end{equation}
Here $a_p$ and $M_p$ are the planet's semimajor axis and mass, and $M_*$ is the stellar mass.  With the planetary orbital frequency $\Omega_p^2=GM_*/a_p^3$, Equation~(\ref{eq:hill}) implies
\begin{equation}
GM_p=3\Omega_p^2\RH^3.
\label{eq:hillidentity}
\end{equation}
Write the specific angular momentum delivered to the Hill sphere as
\begin{equation}
j=\lambda\Omega_p\RH^2.
\label{eq:jlambda}
\end{equation}
The circularization radius obtained by conserving $j$ is then
\begin{equation}
r_{\rm c}=\frac{j^2}{GM_p}
=\frac{\lambda^2}{3}\RH.
\label{eq:rclambda}
\end{equation}
Thus $r_{\rm c}$ cannot be obtained from $\RH$ alone.  It depends quadratically on the dimensionless angular momentum $\lambda$ of the gas that is actually entering the Hill sphere.

For gas accreted locally before a deep gap has formed, \citet{Lissauer1995} estimated the mass-flux-weighted angular momentum as
\begin{equation}
\begin{split}
j_{\rm pre}
&\simeq-\Omega_p
\frac{\displaystyle\int_0^{\RH}(3/2)x^3\,dx}
{\displaystyle\int_0^{\RH}x\,dx}
+\Omega_p\RH^2\\
&=\frac{1}{4}\Omega_p\RH^2.
\end{split}
\label{eq:lissauerj}
\end{equation}
Here $x$ is the radial offset from the planet in the local coorbital calculation.  The first term is the Keplerian-shear contribution in the frame rotating with the planet; the second transforms back to the inertial frame.  Substituting $\lambda=1/4$ into Equation~(\ref{eq:rclambda}) gives
\begin{equation}
r_{\rm c,pre}=\frac{\RH}{48}.
\label{eq:rcpre}
\end{equation}

After a gap forms, gas can enter the Hill sphere through a Lagrange region at the gap edge.  For PDS~70~c, gas supplied from the reservoir between the two planets enters through the starward $L_1$ entry region.  If its velocity is small in the rotating frame at entry, the frame contribution dominates \citep{QuillenTrilling1998,MosqueiraEstrada2003a,Estrada2009}, and
\begin{equation}
j_{\rm gap}\sim\Omega_p\RH^2,
\qquad
r_{\rm c,gap}\sim\frac{\RH}{3}.
\label{eq:rcgap}
\end{equation}
The factor-of-four change in $j$ produces a factor-of-sixteen change in $r_{\rm c}$.

For $a_p=34.5$~au, $M_*=0.76\,\Msun$, and $M_p=4$--$12\,\Mjup$,
\begin{align}
\RH&=4.1\text{--}5.9~{\rm au},\\
r_{\rm c,pre}&=0.085\text{--}0.12~{\rm au},\\
r_{\rm c,gap}&=1.4\text{--}2.0~{\rm au}.
\label{eq:rcnumbers}
\end{align}
The fully dust-dominated thermal examples and the observational upper limit span approximately 0.46--1.2~au.  This scale is several times larger than the compact pre-gap circularization radius.

\begin{figure}[t]
\centering
\includegraphics[width=\columnwidth]{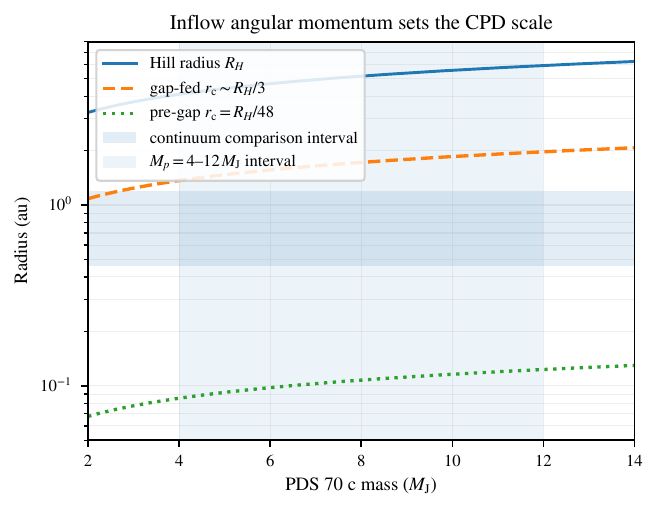}
\caption{Hill and circularization-radius scalings for PDS~70~c.  The pre-gap Lissauer estimate gives $r_{\rm c}=\RH/48$, whereas low-relative-velocity inflow through the starward $L_1$ entry region of a developed gap gives $r_{\rm c}\sim\RH/3$.  The shaded horizontal interval runs from the approximately 0.46-au radius obtained for the fully dust-dominated Benisty-normalized radial temperature profile to the 1.2-au ALMA upper limit.  The lower edge is a conditional thermal-model result, not a model-independent observational bound.  The observed stage is a common-gap stage, not the pre-gap stage used to obtain $\RH/48$.}
\label{fig:hill}
\end{figure}

\section{Large-scale Gap Formation}\label{sec:gapdepletion}

\subsection{Conditional clearing of a finite common-gap reservoir}

The PDS~70 planets occupy one common gap, so their final growth and their circumplanetary supply cannot be prescribed independently.  As the planets transfer angular momentum to the surrounding gas, material in the common planetary region is displaced toward the inner and outer circumstellar disks or transferred into the two Hill regions.  We reduce that region to a single finite reservoir: $M_{\rm gap}(t)$ is the total circumstellar gas mass remaining in the common-gap control volume, $A_{\rm gap}$ is its fixed planar area, and $\Sigma_{\rm gap}(t)=M_{\rm gap}(t)/A_{\rm gap}$ is its mean surface density.

The time-dependent surface-density and wave-deposition problem is classical \citep{GoldreichTremaine1980,Rafikov2002Gap,Crida2006,CordwellRafikov2024}.  We use the standard one-sided tidal-torque form
\begin{equation}
|\mathcal{T}_{p,\pm}|=
C_Tq_p^2\Sigma_{\rm gap}a_p^4\Omega_p^2h_p^{-3},
\label{eq:one_sided_torque}
\end{equation}
where $p\in\{b,c\}$ labels the planet, $\mathcal{T}_{p,+}$ and $\mathcal{T}_{p,-}$ label the one-sided torques on the outer and inner sides of its orbit, respectively, $M_p$ and $a_p$ are its mass and semimajor axis, $M_*$ is the stellar mass, $\Omega_p=(GM_*/a_p^3)^{1/2}$ is the planet's orbital frequency, and $H_p$ is the local circumstellar-disk pressure scale height.  Only the two boundary-directed lobes remove gas from the common-gap control volume: $\mathcal{T}_{b,-}$ sends material toward the inner disk and $\mathcal{T}_{c,+}$ sends material toward the outer disk.  The opposite lobes redistribute gas within the control volume and are not counted as additional reservoir-loss channels.  The dimensionless quantities $q_p=M_p/M_*$ and $h_p=H_p/a_p$ are the planet-to-star mass ratio and disk aspect ratio, and $C_T$ is the adopted dimensionless torque coefficient.  We assign planet $p$ a radial clearing half-width $w_p=\eta_wR_{{\rm H},p}$, where $R_{{\rm H},p}$ is its Hill radius from Equation~(\ref{eq:hill}) and $\eta_w$ is the dimensionless control-volume width.  Moving gas across this distance requires the change in Keplerian specific angular momentum
\begin{equation}
\Delta\ell_p\simeq
\left|\frac{d\ell}{dr}\right|_{a_p}w_p
=\frac{1}{2}a_p\Omega_pw_p.
\end{equation}
Here $\ell(r)=(GM_*r)^{1/2}$ and $\Delta\ell_p$ is evaluated across $w_p$ at the orbit of planet $p$.  This is the usual angular-momentum clearing estimate, $\dot M=|\mathcal{T}|/\Delta\ell$ \citep{PaardekooperJohansen2018,DipierroLaibe2017}.  We adopt $\eta_w=2.5$, the rounded nonlinear horseshoe half-width found by \citet{MassetDAngeloKley2006}; it is a bookkeeping width rather than a measured gap-edge width.

Net Hill-region transfer is represented by
\begin{equation}
\dot M_{H,p}
=C_{H,p}\Sigma_{\rm gap}\Omega_pR_{{\rm H},p}^2,
\label{eq:gap_hill_rate}
\end{equation}
the standard Hill rate \citep{Rosenthal2020,Popovas2018,ChoksiChiangFungZhu2023}.  Here $\dot M_{H,p}$ is the net rate at which mass leaves the common-gap reservoir through the Hill control surface of planet $p$, and the dimensionless coefficient $C_{H,p}$ includes the unresolved difference between gross inflow and recycling outflow.  It is not the accretion rate onto the planet or the retained CPD rate.  Three-dimensional gap-flow simulations are required to calibrate $C_{H,p}$ and the true mass supply; however, specific angular momentum can be estimated separately from the ballistic calculation in Section~\ref{sec:ballistic}.

\subsection{Finite-reservoir clearing timescale for Jupiter--Saturn and PDS~70~b--c}

For the one-zone comparison, $a_b<a_c$, and the control-volume area and its mean surface density introduced above are
\begin{align}
\Sigma_{\rm gap}&=\frac{M_{\rm gap}}{A_{\rm gap}},\\
A_{\rm gap}&=\pi\!\left(a_c+\eta_wR_{{\rm H},c}\right)^2
-\pi\!\left(a_b-\eta_wR_{{\rm H},b}\right)^2.
\end{align}
Thus $A_{\rm gap}$ is the area of the annulus between the inner boundary $a_b-\eta_wR_{{\rm H},b}$ and the outer boundary $a_c+\eta_wR_{{\rm H},c}$.  Equations~(\ref{eq:one_sided_torque}) and (\ref{eq:gap_hill_rate}) are then linear in $M_{\rm gap}$.  Define
\begin{align}
\gamma_{T,p}
&=\frac{C_Tq_p^2a_p^4\Omega_p^2h_p^{-3}}
{A_{\rm gap}\Delta\ell_p},&
\gamma_{H,p}
&=\frac{C_{H,p}\Omega_pR_{{\rm H},p}^2}{A_{\rm gap}},\\
\Gamma_T&=\gamma_{T,b}+\gamma_{T,c},&
\Gamma&=\Gamma_T+\gamma_{H,b}+\gamma_{H,c}.
\label{eq:gap_gamma_components}
\end{align}
The coefficients $\gamma_{T,b}$ and $\gamma_{T,c}$ represent the boundary-directed $\mathcal{T}_{b,-}$ and $\mathcal{T}_{c,+}$ lobes, while $\gamma_{H,p}$ represents net Hill-region transfer by planet $p$; all have dimensions of inverse time.  Their torque-only sum is $\Gamma_T$, while $\Gamma$ is the sum over all four reservoir-loss channels.  The four linear channels give $dM_{\rm gap}/dt=-\Gamma M_{\rm gap}$.  Writing $M_{\rm gap,0}\equiv M_{\rm gap}(0)$ for the initial reservoir mass, the solution is
\begin{equation}
M_{\rm gap}(t)
{}=M_{\rm gap,0}e^{-\Gamma t}
=M_{\rm gap,0}e^{-t/t_{\rm dep}},
\quad t_{\rm dep}\equiv\Gamma^{-1}.
\label{eq:gap_exp_solution}
\end{equation}
\begin{equation}
\dot M_{H,p}(t)
{}=\gamma_{H,p}M_{\rm gap,0}e^{-t/t_{\rm dep}}.
\label{eq:gap_hill_solution}
\end{equation}
For each destination $i$---the inner disk, outer disk, Hill region of b, or Hill region of c---let $\gamma_i$ be the corresponding coefficient above and $M_i(t)$ the cumulative mass delivered there, with $M_i(0)=0$.  Then
\begin{align}
M_i(t)&=\frac{\gamma_i}{\Gamma}M_{\rm gap,0}
\left(1-e^{-t/t_{\rm dep}}\right),\\
M_{\rm gap}(t)+\sum_iM_i(t)&=M_{\rm gap,0}.
\label{eq:gap_destination_mass}
\end{align}

We evaluate the two architectures with the same illustrative closure, $C_T=0.01$, $C_{H,b}=C_{H,c}=0.1$, and $\eta_w=2.5$.  Instead of assigning one common aspect ratio, we calculate $h_p=[k_{\rm B}T(a_p)a_p/(\mu m_{\rm H}GM_*)]^{1/2}$ from the directly illuminated, optically thin temperature profiles in Figure~\ref{fig:passivetemperature}, where $\mu$ is the mean molecular weight in units of the hydrogen mass $m_{\rm H}$; we adopt $\mu=2.34$.  For the representative 3-Myr Solar profile this gives $h_{\rm J}=0.0495$ and $h_{\rm S}=0.0577$; the corresponding PDS~70 profile gives $h_b=0.0725$ and $h_c=0.0811$.  The Jupiter--Saturn calculation uses $M_*=1\,\Msun$, $(M_b,a_b)=(1.00\,\Mjup,5.20~{\rm au})$, and $(M_c,a_c)=(0.299\,\Mjup,9.58~{\rm au})$.  The PDS~70 calculation uses $M_*=0.76\,\Msun$, $(M_b,a_b)=(5\,\Mjup,22~{\rm au})$, and $(M_c,a_c)=(7.5\,\Mjup,34.5~{\rm au})$.  The adopted b mass is a convenient representative value close to the $4.9\,\Mjup$ dynamical upper limit of \citet{Trevascus2025}; the c mass is an illustrative value within the $4$--$12\,\Mjup$ interval considered by \citet{Shibaike2026}.

The resulting comparison yields
\begin{align}
t_{\rm clear}^{\rm JS}&=\Gamma_T^{-1}
=2.1\times10^4~{\rm yr},\label{eq:gap_clearing_js}\\
t_{\rm dep}^{\rm JS}&=\Gamma^{-1}
=1.2\times10^4~{\rm yr},\\
t_{\rm clear}^{\rm PDS70}&=(0.86\text{--}2.1)\times10^4~{\rm yr},\\
t_{\rm dep}^{\rm PDS70}&=(0.73\text{--}1.6)\times10^4~{\rm yr}.
\label{eq:gap_clearing_comparison}
\end{align}
Here $t_{\rm clear}$ is the reservoir e-folding time if only tidal clearing operates, whereas $t_{\rm dep}$ is the reservoir e-folding time when both tidal clearing and net Hill-region transfer operate.  The superscripts JS and PDS70 denote the Jupiter--Saturn and PDS~70~b--c architectures, respectively.  The PDS~70 intervals are obtained by holding $M_b=5\,\Mjup$ fixed and varying $M_c$ over the cited $4$--$12\,\Mjup$ interval.

A rough mass-budget comparison is also instructive.  We can express the combined planetary mass as a mean gas column over the geometric annulus between the two planetary orbits,
\begin{equation}
\overline\Sigma_{\rm pair}
=\frac{M_{\rm inner}+M_{\rm outer}}
{\pi\left(a_{\rm outer}^2-a_{\rm inner}^2\right)}.
\label{eq:pair_surface_density}
\end{equation}
Here $M_{\rm inner}$ and $M_{\rm outer}$ are the masses of the inner and outer planets, and $a_{\rm inner}$ and $a_{\rm outer}$ are their respective orbital radii.
Using the masses and orbital radii above gives $\overline\Sigma_{\rm pair}=54~{\rm g~cm^{-2}}$ for Jupiter--Saturn.  Holding PDS~70~b at $5\,\Mjup$ and varying c over $4$--$12\,\Mjup$ gives $34$--$65~{\rm g~cm^{-2}}$.

\subsection{Sensitivity to the assumed thermal structure}

The calculation above assumes directly illuminated, optically thin circumstellar temperature profiles for both systems.  Table~\ref{tab:thermal-clearing} shows the result of applying the other matched profiles in Figure~\ref{fig:passivetemperature} while retaining the same torque prescription, control-volume geometry, and coefficients.  We define the pair averages as $\langle h\rangle_{\rm JS}=(h_{\rm J}+h_{\rm S})/2$ and $\langle h\rangle_{\rm PDS}=(h_b+h_c)/2$.  The timescale ratios are computed using the individual aspect ratio at each planet.

\begin{table*}[t]
\centering
\caption{Clearing-time sensitivity to matched circumstellar temperature profiles.  The Solar aspect-ratio ranges span the representative 5- and 3-Myr states; each pair lists the inner and outer planet.  The PDS~70 calculation uses $M_c=7.5\,\Mjup$.}
\label{tab:thermal-clearing}
\footnotesize
\setlength{\tabcolsep}{3.5pt}
\begin{tabular}{lccccc}
\toprule
Thermal profile &
$(h_{\rm J},h_{\rm S})$ &
$(h_b,h_c)$ &
$\langle h\rangle_{\rm PDS70}/\langle h\rangle_{\rm JS}$ &
$t_{\rm clear}^{\rm PDS70}/t_{\rm clear}^{\rm JS}$ &
$t_{\rm dep}^{\rm PDS70}/t_{\rm dep}^{\rm JS}$ \\
\midrule
Direct gray irradiation & $(0.0475\text{--}0.0495,\ 0.0553\text{--}0.0577)$ & $(0.0725,\ 0.0811)$ & 1.43--1.49 & 0.64--0.73 & 0.91--0.98 \\
Passive interior, $H_s/H=2$ & $(0.0284\text{--}0.0298,\ 0.0338\text{--}0.0354)$ & $(0.0460,\ 0.0523)$ & 1.51--1.58 & 0.77--0.89 & 0.85--0.96 \\
Passive interior, $H_s/H=4$ & $(0.0313\text{--}0.0329,\ 0.0373\text{--}0.0391)$ & $(0.0508,\ 0.0577)$ & 1.51--1.58 & 0.77--0.89 & 0.87--0.98 \\
Passive interior, $H_s/H=5$ & $(0.0323\text{--}0.0339,\ 0.0385\text{--}0.0404)$ & $(0.0524,\ 0.0596)$ & 1.51--1.58 & 0.77--0.89 & 0.88--0.99 \\
\bottomrule
\end{tabular}
\end{table*}

Across these matched profiles, the PDS~70 aspect ratios are approximately 1.5 times the corresponding Solar System values.  Their $h_p^{-3}$ dependence suppresses the PDS~70 tidal torques by approximately a factor of three to four and thereby largely offsets the stronger planetary-mass terms.  For the full $M_c=4$--$12\,\Mjup$ range, the matched models give $t_{\rm clear}^{\rm PDS70}/t_{\rm clear}^{\rm JS}\simeq0.4\text{--}1.3$ and $t_{\rm dep}^{\rm PDS70}/t_{\rm dep}^{\rm JS}\simeq0.5\text{--}1.4$.  Thus both planetary systems retain similar gap-clearing times for a range of plausible thermal structures.

\subsection{Connection to the observed chronology}

The calculated timescales measure the clearing time after the common gap has isolated the planetary region.  Once this phase begins, the adopted torque model removes the PDS~70~b--c reservoir on a timescale of order $10^4$~yr rather than millions of years.  The later geometric calculations use the fiducial mass choice $M_c=7.5\,\Mjup$ from within the interval above.  The one-zone torque model establishes that circumplanetary delivery must decline as the reservoir is depleted, but it does not determine when the flow becomes predominantly gap-fed or how much of that flow is retained.

\section{Late-stage Ballistic Assembly of the Circumplanetary Disk}\label{sec:ballistic}

The finite-reservoir calculation establishes that the gas available to the planets declines as the common gap is cleared.  It does not determine when most of the remaining gas enters PDS~70~c through $L_1$ or how much of that gas is retained; those quantities require three-dimensional gap and Hill-sphere simulations.  The ballistic calculation is instead conditional: given a late retained mass $M_{\rm late}$, it asks how much axial angular momentum the adopted $L_1$-fed material carries and where that material circularizes.  The source is the finite reservoir between planets b and c; $L_2$ delivery from the outer circumstellar disk would introduce an additional reservoir that is not part of this calculation.

The Hill equations and the adopted launch prescription are reflection symmetric, so corresponding $L_1$ and $L_2$ ensembles have the same normalized circularization distribution.  Including both sides and renormalizing to the same retained mass would therefore leave the radial result unchanged.  We use $L_1$ alone to maintain consistency with the interplanetary mass source modeled here, not because the normalized kinematic result depends on excluding $L_2$.  The compact pre-gap disk has a distinct scale $r_{\rm c,pre}=\RH/48$ \citep{Lissauer1995,Estrada2009}.

The exact inflow angular momentum is a mass-flux-weighted integral over the three-dimensional Hill boundary.  We estimate it by integrating the three-dimensional Hill equations from a vertical column at the starward $L_1$ entry region, a prescription appropriate to gas supplied from the reservoir between PDS~70~b and c.  The adopted circumstellar aspect ratio is $h_c=H_c/a_c=0.08$, corresponding to approximately 36~K for $\mu=2.34$ and lying within the range represented by the thermal profiles in Figure~\ref{fig:passivetemperature}; for the fiducial parameters, $H_c=2.76$~au and $H_c/\RH=0.546$.  Seventeen launch heights span $-H_c\leq z\leq H_c$, each with $u_z=0$, and the incident density is weighted as $\exp[-z^2/(2H_c^2)]$.  This interval contains 68.3\% of a Gaussian vertical column, and the ensemble is normalized within that central interval rather than over the full vertical flow.  At every height, 289 low-relative-velocity trajectories sample $0.1\leq u_n\leq0.5$ and $|u_t|\leq0.1$.  Here $u_n$, $u_t$, and $u_z$ are the inward-normal, tangential, and vertical rotating-frame entry velocities in units of $\Omega_p\RH$.  The uniform $u_n$--$u_t$ grid defines an illustrative flat velocity-space prior over these intervals; it is not inferred from a global gap-flow calculation.  The full mass-flux weight is proportional to $u_n\exp[-z^2/(2H_c^2)]$.  Of the 4913 launched trajectories, 4205 reach the first crossing of the illustrative planet-centered interaction radius $S_{\rm sh}=0.5\RH$.  At that crossing, orbital energy is assumed to dissipate while the instantaneous planet-centered axial specific angular momentum is retained.  Writing $v_x$ and $v_y$ for the planet-centered rotating-frame velocities at this crossing, the required inertial-frame angular momentum is
\begin{equation*}
j_z=x(v_y+\Omega_px)-y(v_x-\Omega_py)
=xv_y-yv_x+\Omega_p(x^2+y^2).
\end{equation*}
Following the Hill-trajectory bookkeeping used for planetesimal delivery by \citet{EstradaMosqueira2006}, these choices specify an illustrative angular-momentum distribution, not a calculation of the three-dimensional supply, shock, or capture probability.

Define
\begin{equation}
\lambda\equiv\frac{j_z}{\Omega_p\RH^2},
\qquad
\frac{R_{\rm c}}{\RH}=\frac{\lambda^2}{3}.
\label{eq:ballistic_mapping}
\end{equation}
Here $R_{\rm c}$ denotes the circularization radius of an individual sampled trajectory, whereas $r_{\rm c}$ denotes the characteristic or phase-dependent radius used elsewhere.
The flux-weighted ensemble gives
\begin{equation}
\begin{aligned}
\langle\lambda\rangle&\simeq0.83,
&\sigma_\lambda&\simeq0.06,\\
\left\langle R_{\rm c}/\RH\right\rangle&\simeq0.23,
&\sigma_{R_{\rm c}/\RH}&\simeq0.03.
\end{aligned}
\label{eq:ballistic_result}
\end{equation}
Here $\sigma_\lambda$ and $\sigma_{R_{\rm c}/\RH}$ are the corresponding flux-weighted standard deviations.  These displayed values are rounded to reflect the illustrative nature of the ensemble; the unrounded values used for the numerical closure check are given in Appendix~\ref{app:figures}.
For the fiducial PDS~70~c value $\RH=5.1$~au, the full sampled range of circularization radii is approximately 0.5--1.5~au, with a flux-weighted mean of 1.2~au.  The mean deposition area is approximately 120 times that associated with the compact $\RH/48$ pre-gap scale; the analytic zero-entry-velocity limit $R_{\rm c}=\RH/3$ results in a geometric factor of 256.

\begin{figure}[t]
\centering
\includegraphics[width=\columnwidth]{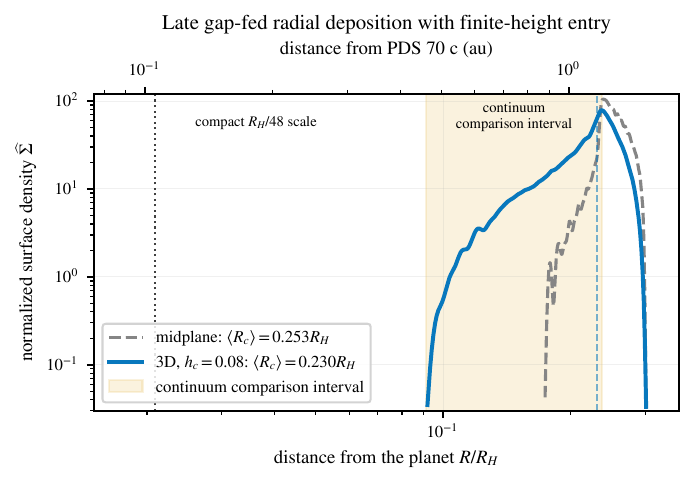}
\caption{Normalized radial deposition profile from the late gap-fed ballistic ensemble.  The solid blue curve is the adopted $L_1$-only, vertically stratified calculation with $h_c=0.08$, corresponding to approximately 36~K and lying within the thermal-profile range shown in Figure~\ref{fig:passivetemperature}; the dashed gray curve shows the independently normalized earlier midplane result.  The dotted line marks the compact $\RH/48$ pre-gap scale, and the blue dashed line marks the finite-height flux-weighted mean circularization radius.  The shaded interval extends from the approximately 0.46-au fully dust-dominated radial-temperature example to the 1.2-au observational upper limit; its lower edge is model dependent.  The ordinate is independent of $M_{\rm late}$; an externally specified retained mass converts it to a physical surface density through Equation~(\ref{eq:ballistic_surface_density}).  The calculation does not determine the late mass supply, retention, or post-deposition redistribution.}
\label{fig:gap-cpd-loading}
\end{figure}

Let $p_x(x)$ be the normalized distribution of circularization radii, with $x=R/\RH=\lambda^2/3$.  Given an independently specified late retained mass $M_{\rm late}$, depositing that mass without subsequent redistribution gives
\begin{equation}
\begin{aligned}
\Sigma_{\rm late}(R\mid M_{\rm late})
&=\frac{M_{\rm late}}{2\pi\RH^2}
\frac{p_x(x)}{x},
\\
\widehat{\Sigma}(x)
&\equiv\frac{2\pi\RH^2\Sigma_{\rm late}}{M_{\rm late}}
=\frac{p_x(x)}{x}.
\end{aligned}
\label{eq:ballistic_surface_density}
\end{equation}
Figure~\ref{fig:gap-cpd-loading} plots the dimensionless profile $\widehat{\Sigma}$, which is independent of $M_{\rm late}$.  Any externally specified $M_{\rm late}$ rescales the ordinate through Equation~(\ref{eq:ballistic_surface_density}) without changing the kinematic radial distribution.  Direct radial integration gives $\int x\widehat{\Sigma}\,dx=1$ and recovers the independently calculated specific-angular-momentum moment $\int\sqrt{3x}\,x\widehat{\Sigma}\,dx=\langle\lambda\rangle\simeq0.83$.

The radial result provides the most direct continuum comparison.  The finite-height ballistic interval, approximately 0.5--1.5~au, overlaps the approximately 0.46--1.2~au scale spanning the fully dust-dominated radial-temperature example and the observational upper limit for PDS~70~c.  The lower edge is model dependent.  Relative to the midplane calculation, vertical averaging broadens the distribution toward smaller radii and shifts the mean from $0.25\RH$ to $0.23\RH$.  Material remains outside this interval, as expected because the model has neither a calculated capture probability nor a treatment of post-impact redistribution.

Stellar tidal truncation provides a separate outer-boundary constraint and generally occurs beyond this mass-weighted deposition radius.  Periodic-orbit calculations place the limiting radius of a cold prograde circumplanetary disk near $0.4\RH$ \citep{MartinLubow2011}, while radiation-hydrodynamic calculations of disks around high-mass protoplanets find radial extents close to $\RH/3$ \citep{AyliffeBate2009}.  Both scales exceed the adopted mean deposition radius, $0.23\RH$, and lie at or beyond the upper end of the sampled distribution, $0.30\RH$.  Thus tidal truncation does not set the mean circularization scale in Figure~\ref{fig:gap-cpd-loading}, although stellar tides may redistribute or remove material in its tail.

\subsection{Hill entry, retention, and subsequent forcing}\label{sec:hill-recycling}

A streamline crossing the Hill surface can horseshoe or recycle outward unless it loses sufficient energy and angular momentum; Hill entry is therefore not capture.  Declining ambient density removes the open recycling flow but does not itself provide the torque required to remove gas already captured into a rotationally supported disk \citep{FungZhuChiang2019,ChoksiChiangFungZhu2023}.  Three-dimensional gap-flow simulations are needed to determine when the late flow becomes predominantly $L_1$-fed, the Hill-sphere mass flux, the recycling--capture balance, the retained mass $M_{\rm late}$, and the vertical deposition and dissipation efficiencies.  They are not needed for the conditional kinematic mapping used here: once the adopted finite-height entry conditions specify $j_z$, angular-momentum conservation gives $r_{\rm c}=j_z^2/(GM_p)$ independently of the retained-mass normalization.

For a specified forcing history, the phase-dependent circularization radius sets the deposition area, $A_{\rm dep}(t)=\pi r_{\rm c}(t)^2$.  The corresponding loading is $\dot\Sigma_{\rm src}=\dot M_{\rm src}/A_{\rm dep}$, and the mechanical power available to drive non-Keplerian motion is $\dot E_{\rm src}=\epsilon_{\rm in}\dot M_{\rm src}v_{\rm rel}^2/2$, where $\epsilon_{\rm in}$ is an effective conversion efficiency that includes unresolved capture and mechanical coupling.  These are conditional inputs, not measurements of CPD-boundary inflow.  Both quantities vanish as the finite source is exhausted; the retained circumplanetary disk requires a separate physical process for gas dispersal.

\section{Transient Inflow-driven Fluctuations}\label{sec:les}

While the observational evidence favors quiescent planet-forming disks, gas inflow can nevertheless drive turbulent fluctuations in a circumplanetary disk by supplying kinetic energy.  This forcing is transient and decays as the inflow declines; it is not evidence for sustained turbulence in an unforced Keplerian disk.  Appendix~\ref{app:les} gives the fluctuation-energy, Reynolds-stress, ram-pressure, and decay calculations.

The stream--disk interaction acts through shocks, momentum loading, and velocity mismatch, but its conversion efficiency, anisotropy, and vertical deposition require a three-dimensional inflow calculation.  Ram pressure can perturb a tenuous surface layer even when the inflow supplies too little energy to stir the full disk column; Figure~\ref{fig:infall-depth} keeps those two limits separate.  The plotted H$\alpha$-derived planetary accretion rates serve only as an empirical comparison and determine neither the CPD-boundary inflow nor the mechanically coupled power.  The reduced calculation adopts a prescribed declining source motivated by depletion of the finite common-gap reservoir.  As that source disappears, the affected layer retreats toward the surface and the fluctuations decay at every height, as shown in Figures~\ref{fig:infall-depth} and \ref{fig:forced-decay} and as derived in Appendix~\ref{app:les}.

\section{Implications for Regular-satellite Formation}\label{sec:satellites}

\subsection{The \SEMM{} model and the PDS~70 comparison}

The \SEMM{} model describes regular-satellite formation in an extended, quiescent circumplanetary disk with inner and outer satellite-forming regions \citep{MosqueiraEstrada2003a,MosqueiraEstrada2003b,Estrada2009,Mosqueira2010SSR}.  Its solids constraint is the mass contained in the satellite-forming disk.  Most of that solid material is ultimately incorporated into the regular satellites, so their masses constrain the solids inventory that the disk must have contained.  The gas content is then reduced relative to that solids inventory so that satellite survival, thermal constraints, and the observed compositional structure can be satisfied.  ``Solids-enhanced'' means that the satellite-forming disk has a larger solids-to-gas ratio than an unprocessed solar-composition mixture.

\subsection{Satellite assembly, disk clearing, and the age of PDS~70}\label{sec:times}

In the \SEMM{} model, the satellite-formation timescale is not set by a local accretion-rate formula.  It is tied to the time required for gas drag to clear the circumplanetary disk of the satellitesimal population available for satellite growth.  For the modeled circumjovian and circum-Saturnian disks, this argument gives formation timescales of $\sim10^6$~yr for Callisto and $\sim10^7$~yr for Iapetus \citep{MosqueiraEstrada2003a,MosqueiraEstrada2003b,Estrada2009,Mosqueira2010SSR}.  These values are outputs of the gas-drag clearing calculation, not evaluations of a local accretion formula.

The $5.4\pm1.0$~Myr age of PDS~70 lies between the Callisto and Iapetus formation timescales.  The secure continuum source at c and the absence of a comparably secure source at b consequently favor an interpretation in which the inner circumplanetary reservoir has been more substantially processed or depleted than that at c while the outer circumplanetary disk and circumstellar reservoir remain active.  However, our satellite-formation model must be modified for the PDS~70 system because these planets are more massive and their Hill radii are larger.

For the fiducial PDS~70 parameters, Equation~(\ref{eq:hill}) gives
\begin{equation}
R_{{\rm H},b}=2.81~{\rm au},\qquad
R_{{\rm H},c}=4.1\text{--}5.9~{\rm au},
\label{eq:pds70_hill_comparison}
\end{equation}
where the interval for c corresponds to $M_c=4$--$12\,\Mjup$.  Its Hill radius is therefore $1.46$--$2.10$ times that of b; a geometric area proportional to $\RH^2$ would be larger by a factor of approximately $2.1$--$4.4$.  At a common fractional Hill radius $r=f\RH$, the local orbital period is
\begin{equation}
P_{\rm sat}(f\RH)
=2\pi\left[\frac{(f\RH)^3}{GM_p}\right]^{1/2}
=\frac{f^{3/2}}{\sqrt{3}}P_p,
\label{eq:hill_period_scaling}
\end{equation}
where $P_p$ is the planet's orbital period.  Thus $P_{{\rm sat},c}/P_{{\rm sat},b}=(a_c/a_b)^{3/2}=1.96$ at the same $f$.

To compare the four giant planets with one consistent calculation, we spread one Callisto mass of solids uniformly inside $R_{{\rm H},p}/3$ around Jupiter, Saturn, PDS~70~b, and PDS~70~c, and adopt a gas-to-solids ratio of 100 in every case.  Thus
\begin{equation}
\Sigma_s=\frac{M_{\rm Callisto}}{\pi(R_{{\rm H},p}/3)^2},
\qquad
\Sigma_g=100\Sigma_s.
\label{eq:controlled_satellite_columns}
\end{equation}
The isolation mass, physical radius, and embryo Hill radius are
\begin{equation}
\begin{aligned}
M_{\rm iso}&=\left[
\frac{2\pi b_{\rm iso}r^2\Sigma_s}{(3M_p)^{1/3}}
\right]^{3/2},\\
s_{\rm iso}&=\left(\frac{3M_{\rm iso}}{4\pi\rho_s}\right)^{1/3},\\
R_{{\rm H},{\rm iso}}&=r\left(\frac{M_{\rm iso}}{3M_p}\right)^{1/3}.
\end{aligned}
\label{eq:isolation_properties}
\end{equation}
We choose a feeding-zone half-width of $2.5R_{{\rm H},{\rm iso}}$ on each side, so the full-width coefficient in Equation~(\ref{eq:isolation_properties}) is $b_{\rm iso}=5$, and take $\rho_s=1~{\rm g~cm^{-3}}$.\footnote{For a compact rock--ice mixture, $\rho_s\simeq1.5~{\rm g~cm^{-3}}$ is more appropriate; however, this distinction does not affect the order-of-magnitude comparison made here.}

For the drag estimate, we use the representative outer-disk temperatures $(T_{\rm J},T_{\rm S},T_b,T_c)=(130,90,45,36)$~K shown in Figure~\ref{fig:passivetemperature}, mean molecular weight $\mu=2.34$, a vertically Gaussian gas column, $\rho_g=\Sigma_g/(\sqrt{2\pi}H)$, a local pressure slope $d\ln P/d\ln r=-3/2$, and quadratic high-Reynolds-number drag with $C_D=0.44$.  With
\begin{equation}
\begin{aligned}
\eta&=-\frac{1}{2}\left(\frac{H}{r}\right)^2\frac{d\ln P}{d\ln r},\\
|v_{r,{\rm drag}}|&\simeq
\frac{3C_D\rho_g}{4\rho_s s_{\rm iso}}\eta^2v_Kr,\\
t_{{\rm drag},{\rm orb}}&=
\frac{r}{|v_{r,{\rm drag}}|},\\
t_{{\rm drag},{\rm mig}}&=
\frac{R_{{\rm H},{\rm iso}}}{r}t_{{\rm drag},{\rm orb}}.
\end{aligned}
\label{eq:isolation_encounter_time}
\end{equation}
Here $\eta$ is the pressure-support parameter, $H$ is the local gas pressure scale height, and $v_K$ is the circumplanetary Keplerian speed.  The corresponding linear, three-dimensional, locally isothermal Type-I migration time is \citep{Tanaka2002}
\begin{equation}
\begin{aligned}
t_{I,{\rm orb}}&=C_I\frac{\Omega r^2c_s^2}
{\Sigma_gG^2M_{\rm iso}},\\
C_I&=\frac{1}{2.7+1.1\beta},
\qquad \Sigma_g\propto r^{-\beta},\\
t_{I,{\rm mig}}&=\frac{R_{{\rm H},{\rm iso}}}{r}t_{I,{\rm orb}}.
\end{aligned}
\label{eq:isolation_torque_time}
\end{equation}
Here $\Omega=(GM_p/r^3)^{1/2}$ is the local circumplanetary orbital frequency, $c_s=H\Omega$ is the isothermal sound speed, and $t_{I,{\rm orb}}=r/|\dot r_I|$ is the local Type-I orbital-decay time.  The orbital-decay times $t_{{\rm drag},{\rm orb}}$ and $t_{I,{\rm orb}}$ test whether an embryo can migrate a radial distance of the order of its orbit.  The Hill-radius migration times $t_{{\rm drag},{\rm mig}}$ and $t_{I,{\rm mig}}$ measure the timescales for migration across the embryo's Hill radius and test whether the embryo--gas interaction can result in the persistent collisional evolution of otherwise isolated embryos.  The uniform-column calculation has $\beta=0$ and $C_I=0.37$.  We evaluate all the dynamical quantities locally at $r=R_{{\rm H},p}/10$.  The resulting values are listed in Table~\ref{tab:controlled-embryo-drift}.\footnote{The SEMM model yields somewhat different values for several reasons: first, because it uses $g/s = 10$; second, because dynamical quantities are evaluated at the location of the satellites; and third, because it uses the outer satellites Callisto and Iapetus to determine the mass of solids for Jupiter and Saturn, respectively.}

\begin{table*}[t]
\centering
\caption{Satellite-embryo drift comparison.  Every row uses one Callisto mass of solids uniformly inside $R_{{\rm H},p}/3$, gas-to-solids ratio 100, and local evaluation at $r=R_{{\rm H},p}/10$.  The orbital-decay times apply to migration over a distance of order $r$; the Hill-radius migration times apply to migration across an embryo's Hill radius.}
\label{tab:controlled-embryo-drift}
\scriptsize
\setlength{\tabcolsep}{4.0pt}
\renewcommand{\arraystretch}{1.12}
\textit{(a) Subnebula parameters and satellite embryo properties}\\[2pt]
\begin{tabular}{@{}lrrrrrr@{}}
\toprule
Planet & $T$ (K) & $R_{{\rm H},p}$ (au) & $\Sigma_s$ (${\rm g~cm^{-2}}$) & $M_{\rm iso}$ ($10^{23}$~g) & $s_{\rm iso}$ (km) & $R_{{\rm H},{\rm iso}}$ ($10^{-4}$~au) \\
\midrule
Jupiter   & 130 & 0.355 & 10.9  & 3.99 & 457 & 1.46 \\
Saturn    &  90 & 0.437 & 7.20  & 7.30 & 559 & 3.30 \\
PDS~70~b &  45 & 2.81  & 0.174 & 1.79 & 349 & 5.18 \\
PDS~70~c &  36 & 5.05  & 0.054 & 1.46 & 327 & 7.60 \\
\bottomrule
\end{tabular}
\par\medskip
\textit{(b) Orbital-decay and Hill-radius migration times}\\[2pt]
\begin{tabular}{@{}lrrrr@{}}
\toprule
Planet & $t_{{\rm drag},{\rm orb}}$ (yr) & $t_{I,{\rm orb}}$ (yr) & $t_{{\rm drag},{\rm mig}}$ (yr) & $t_{I,{\rm mig}}$ (yr) \\
\midrule
Jupiter   & $7.3\times10^6$  & $7.2\times10^6$  & $3.0\times10^4$ & $3.0\times10^4$ \\
Saturn    & $7.0\times10^6$  & $2.5\times10^6$  & $5.3\times10^4$ & $1.9\times10^4$ \\
PDS~70~b & $8.6\times10^9$  & $2.2\times10^9$  & $1.6\times10^7$ & $4.0\times10^6$ \\
PDS~70~c & $5.4\times10^{10}$ & $1.1\times10^{10}$ & $8.2\times10^7$ & $1.7\times10^7$ \\
\bottomrule
\end{tabular}
\end{table*}

Since the orbital-decay times are approximately $2\times10^9$~yr at PDS~70~b and $10^{10}$~yr at c, the isolation-mass embryos are stranded in the outer disks of both PDS planets.  This is unlike the Jupiter and Saturn disks, where orbital decay takes place on the gas-dissipation timescale of $\sim10^7$~yr.  This separation of timescales is a direct consequence of the larger Hill radii and lower circumplanetary gas densities in the PDS~70 system.

On the other hand, the embryo migration times across their Hill radii, $\sim4\times10^6$~yr at b and $\sim2\times10^7$~yr at c, indicate that detailed modeling of the persistent migration-driven embryo collisional evolution could explain the presence of a circumplanetary dust signature at c but a weaker signal at b, provided that gas dissipation occurs on a faster timescale at b than at c.  However, for this argument to go through, we would need either observational or theoretical constraints on the gas columns in the circumplanetary disks of the planets.  Alternatively, continued delivery of planetesimal fragments from the outer PDS~70 circumstellar disk may contribute to the dust observed around c but not b (see Section~\ref{sec:fragments}).

\subsection{The solids mass and the Callisto comparison}

Callisto has a mass of $0.0180\,\Mearth$.  The discovery-analysis dust interval gives
\begin{equation}
\frac{M_{\rm d,c}}{M_{\rm Callisto}}
=0.4\text{--}1.7
\quad\text{for}\quad
M_{\rm d,c}=0.007\text{--}0.031\,\Mearth.
\label{eq:callistoratio}
\end{equation}
The broader multi-epoch interval spans about 0.4--3.5 Callisto masses.  Dust already incorporated into larger bodies is not counted by the continuum.  Thus the inferred mass in radiating grains alone is of the order needed to form a major regular satellite.

In the optically thick branch, the continuum no longer measures the total mass, but it still requires an extended high-column photosphere.  Optical thickness can hide additional solids; it cannot turn the detected component into a solids-depleted reservoir.

\subsection{The radial scale and the common gap}

The approximately 0.46--1.2~au scale spanning the fully dust-dominated radial-temperature example and the observational upper limit corresponds to $0.08$--$0.29\,\RH$ over the adopted planet-mass range.  The lower edge is model dependent, but these scales are much larger than the pre-gap circularization radius, $\RH/48$, and lie on the scale expected when gas is delivered through a developed gap.  This is the geometry used in the extended \SEMM{} construction: a compact inner disk and an outer satellite-forming region.

The radial scale also enters the satellitesimal-clearing history.  The local circumplanetary orbital period increases as $r^{3/2}$, so an extended solids reservoir includes an outer region with longer dynamical, drag-clearing, and aggregation times.  Compressing the reservoir into a disk of order $\RH/48$ would remove the radial domain responsible for the prolonged satellite-assembly history.

\subsection{What the SR~12~c observations require}

SR~12~c provides a model-dependent solids inventory together with independent evidence that gas remains in its circumplanetary environment, supporting a gas- and solids-bearing satellite-forming disk.  Because the continuum source is unresolved, it does not provide a measured radial scale.  The observations also do not identify its supply history; the gap-controlled inflow calculation developed here applies specifically to the PDS~70~b--c common-gap architecture.

\subsection{Gas-to-dust uncertainty and planetesimal-fragment delivery}\label{sec:fragments}

The CPD gas mass is not presently known well enough to assign a reliable gas-to-dust ratio.  This leaves open the planetesimal collisional model developed by \citet{EstradaMosqueira2006}, in which a large fraction of the circumplanetary mass resides in solids and the observable dust is continually replenished by fragments generated in a giant-planet-induced collisional cascade.  In this interpretation, collisions among material already retained in the CPD replenish small grains; PDS~70~b may have incorporated or dynamically isolated a larger fraction of its solids, reducing its collision and dust-production rates.  A separate possibility is that planetesimals from the outer circumstellar disk continue to reach c but not b.  The current continuum measurements do not distinguish internal dust production from external planetesimal delivery.

The observable grain population can be supplied by direct dust or pebble inflow, collisional fragmentation, and ablation of disk-crossing bodies, and depleted through incorporation into satellitesimals and satellites, capture by the planet, or ejection.  The continuum constrains only grains that contribute appreciably to the opacity; it does not directly count kilometer-scale planetesimals or already formed moons.

In the \SEMM{} model, ablation of icy and rocky planetesimal fragments crossing the circumplanetary gas is a viable mechanism for enriching the satellite-forming disk in solids \citep{MosqueiraEstrada2003a,Mosqueira2010SSR}.  It also supplies the leading explanation for the ice-rich composition of Iapetus relative to other regular satellites and outer Solar System objects \citep{Mosqueira2010Iapetus}.

The two limiting continuum interpretations lead to the same conclusion.  If the emission is optically thin, it measures an instantaneous grain mass comparable to a major regular satellite.  If it is optically thick, it identifies a high-column region where fragmentation, drift, concentration, and satellitesimal formation can cycle mass through the observable grain sizes.  In either case, the observed dust is a tracer of an active solids-processing reservoir.

\section{Conclusions}\label{sec:conclusions}

The PDS~70 system connects two forming giant planets, a common large-scale gap, and a directly observed circumplanetary solids reservoir; SR~12~c supplies the second bound planetary-mass disk.  Together they constrain giant-planet growth and satellite formation.

\begin{enumerate}
\item The compact continuum source at PDS~70~c is a satellite-forming circumplanetary reservoir.  It is spatially associated with the accreting planet, lies inside its Hill sphere, and has been recovered at multiple ALMA frequencies and epochs.

\item In the optically thin limit, the 855-$\mu$m PDS~70~c flux implies $0.007$--$0.031\,\Mearth$ of dust for the discovery-analysis opacities at 26~K, comparable to Callisto.  A fully dust-dominated, uniform, isothermal optically thick emitter has equivalent coplanar radii of 0.58 and 0.66~au at 26 and 22~K, respectively, while the Benisty-normalized radial temperature profile gives approximately 0.46~au for a negligible inner radius.  A non-dust contribution reduces these radii; for the isothermal examples, $f_{\rm d}=0.5$ gives approximately 0.41--0.46~au.  The observational upper limit is about 1.2~au.

\item SR~12~c retains a gas- and solids-bearing circumplanetary disk even though its current mass-growth time is $(1.9\pm0.6)\times10^9$~yr and continued accretion for another Myr would add only $(5.2\pm1.7)\times10^{-4}$ of its present mass.  Planetary assembly and circumplanetary-disk survival therefore operate on distinct timescales.

\item A common-gap reservoir-clearing calculation shows that comparing Jupiter--Saturn with PDS~70~b--c is informative.  Expressing the combined planetary mass in each system as a mass-equivalent mean gas surface density over the annulus between the two planets gives similar values, approximately $50~{\rm g~cm^{-2}}$ for the Jupiter--Saturn system and for the fiducial PDS~70 pair.  Applying the same torque-clearing prescription to matched thermal structures likewise gives comparable gap-clearing times of order $10^4$~yr for both systems.

\item The PDS~70~c CPD radial scale can be explained by the angular momentum of the inflow.  For $j=\lambda\Omega_p\RH^2$, $r_{\rm c}=\lambda^2\RH/3$.  The zero-entry-velocity estimate for inflow through the starward $L_1$ entry region is $r_{\rm c}\sim\RH/3$, or $1.4$--$2.0$~au, on the same au scale as the continuum constraints and much larger than the compact $\RH/48$ pre-gap circularization radius.  The finite-height ensemble with nonzero entry velocities spans approximately 0.5--1.5~au and overlaps the approximately 0.46--1.2~au emitting scale.

\item The estimated radial satellite embryo clearing times for the PDS~70~b--c outer circumplanetary disks are of order $10^{9}$--$10^{10}$~yr.  Therefore, isolation-mass satellite embryos are stranded in the outer disks of b and c.  This differs from the outcome for the smaller circumjovian and circum-Saturnian disks, where migration can clear embryos on a timescale of $\sim10^7$~yr, comparable to the timescale over which the circumplanetary gas dissipates.

\item Given that the time for embryo migration across its own Hill radius is $\sim10^7$~yr at PDS~70~c, detailed modeling work will be needed to determine whether a decline in the embryo collision rate and a concomitant decline in collisional dust production can be linked to the absence of a comparably secure circumplanetary continuum detection at PDS~70~b.  But we stress that such a study would ideally include both a quantitative model of gas dissipation at planet b compared to c and observational constraints on the circumplanetary gas columns.  Alternatively, continued planetesimal delivery from the outer PDS~70 circumstellar disk to c but not b remains a viable explanation for the difference in continuum emission between the two PDS planets.

\item The satellite systems cannot be separated from the two-planet gap.  Gap formation limits further planetary growth, changes the specific angular momentum and deposition radius of CPD inflow, and ultimately exhausts circumplanetary supply from the finite interplanetary reservoir in the isolated limit.  The gas between and in the vicinity of the two giant planets is a finite circumstellar reservoir.  Planetary tidal torques partition this gas among the inner disk, the outer disk, and the two Hill regions.

\end{enumerate}

Quiescence is not a cosmetic feature of our model: introducing persistent global $\alpha$-transport would remove the model's predictive closure.  Such transport would connect the finite common-gap reservoir to an external supply and would leave the final planetary masses unbounded unless an additional cutoff were imposed.  It would also introduce migration, circumplanetary-disk drainage, leakage across the gap, and possible replenishment of the circumplanetary disk around PDS~70~b while creating conditions less favorable to planetesimal and satellitesimal formation \citep{Mosqueira2022LPSC}.  Additional assumptions would then be required to recover the observed planetary masses, preserve the secure continuum source at c relative to the much weaker and presently tentative emission at b, explain the survival of the gas- and solids-bearing disk around SR~12~c after planetary growth is effectively complete, and retain the favorable satellite-formation conditions advanced here.  Nevertheless, the lack of an observational constraint on the gas-to-dust ratio means that a solids-dominated planetesimal model should also be considered \citep{EstradaMosqueira2006}.

\section*{Acknowledgments}

This paper was developed in collaboration with ChatGPT~5.6~Pro and GPT-5.6~Sol (high), with editorial and manuscript-preparation assistance from Claude Opus~5 and GLM~5.3.

\appendix
\renewcommand{\theequation}{\Alph{section}\arabic{equation}}
\renewcommand{\theHequation}{appendix.\Alph{section}.\arabic{equation}}

\section{Inflow-driven Fluctuations: Supporting Calculation}\label{app:les}

The main text uses two results from this reduced calculation: inflow can perturb a surface layer without stirring the full CPD column, and the resulting non-Keplerian fluctuations decay when a finite source ends.  This appendix specifies the calculations used for Figures~\ref{fig:infall-depth} and \ref{fig:forced-decay}.  It is not a resolved model of the three-dimensional stream--disk interaction and does not replace the transient inflow source with a persistent $\alpha$ stress.

\subsection{Energy and vertical-depth scalings}

At CPD radius $R_0$ around a planet of mass $M_p$, let
$\Omega_0=(GM_p/R_0^3)^{1/2}$, $H=c_s/\Omega_0$, and
$q=-d\ln\Omega/d\ln R=3/2$.  For non-Keplerian velocity
fluctuations $u_i'$, define
\begin{equation}
R_{ij}=\langle u_i'u_j'\rangle,
\qquad
K=\frac12(R_{xx}+R_{yy}+R_{zz}).
\end{equation}
The local fluctuation-energy budget is
\begin{equation}
\frac{dK}{dt}
=q\Omega_0R_{xy}+P_{\rm in}-D-\nabla\cdot\bm F_K,
\label{eq:turbulentbudget}
\end{equation}
where $P_{\rm in}$ is the mechanical power converted into fluctuations per unit mass, $D$ is the dissipation rate, and $\bm F_K$ is the spatial flux of fluctuation energy.  Coriolis forces do no scalar work but control the stress correlation $R_{xy}$.  In an unforced Keplerian flow, epicyclic exchange does not maintain the correlation needed for sustained extraction of shear energy, and the fluctuations decay \citep{BalbusHawley1998,Hawley1999}.

\begin{figure*}[t]
\centering
\includegraphics[width=0.90\textwidth]{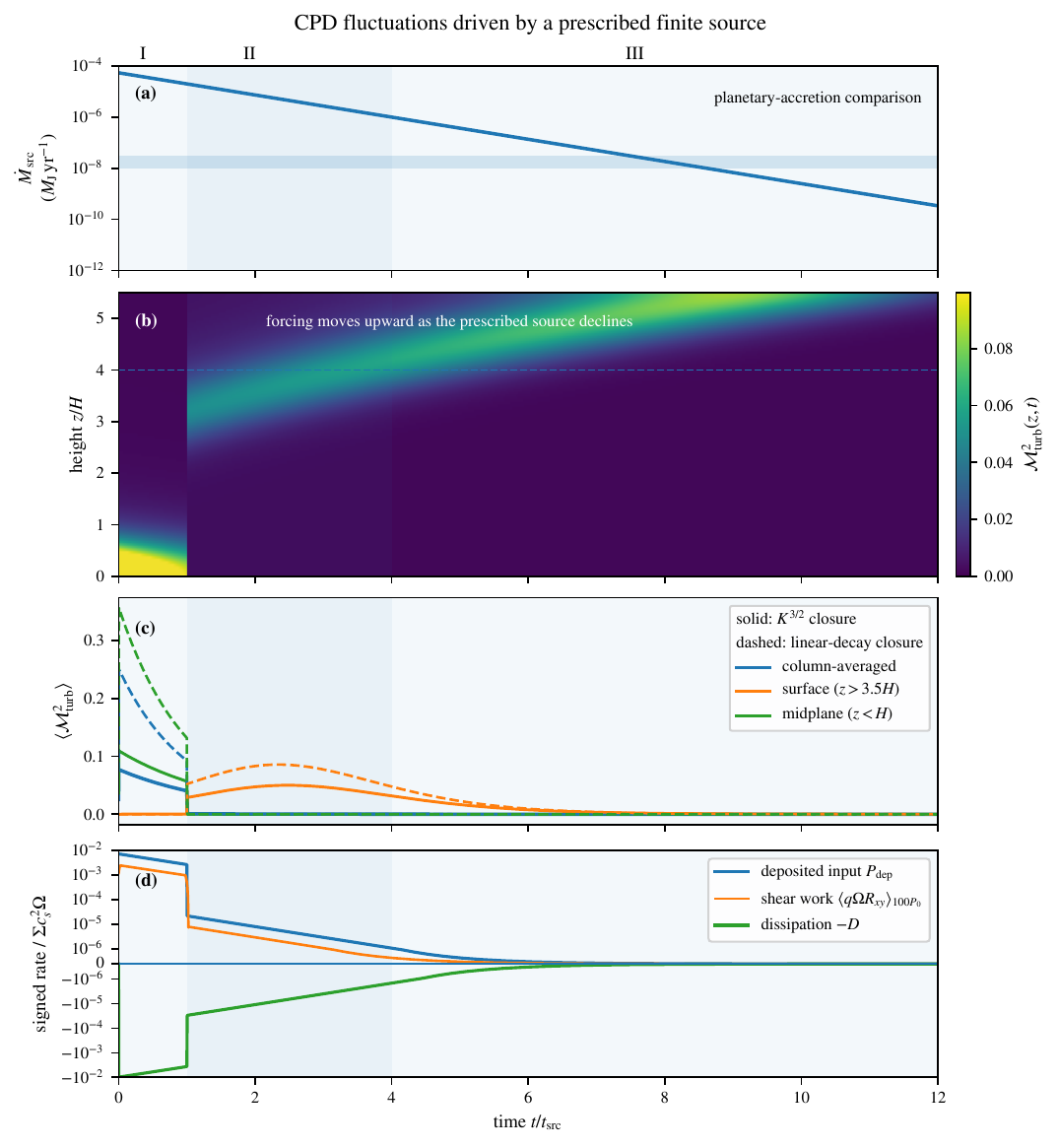}
\caption{Reduced CPD stress response to a prescribed declining source.  Panel (a) shows the adopted source history; the HST-derived planetary accretion band is a comparison only.  Panels (b) and (c) show the vertical response and its density-weighted averages for the nonlinear cascade (solid) and one-orbit linear closure (dashed).  The panel-(c) curves are displayed with a centered $10P_0$ moving average; the integrated initial stress tensor itself is zero.  Panel (d) gives the deposited incident power, shear exchange averaged over $100P_0$, and the nonlinear sink.  At $t/t_{\rm src}=1$, the imposed switch from $\lambda_{\rm ph}=0.25$ to 0.83 increases the deposition radius by $(0.83/0.25)^2\simeq11$ and its area by $(0.83/0.25)^4\simeq122$.  The abrupt transition to surface-layer forcing is therefore produced by this imposed geometric change, superposed on the smoothly declining exponential mass source.  The Roman numerals mark (I) pre-transition whole-column forcing at $t/t_{\rm src}<1$, (II) the surface-layer response after the area expansion at $1\leq t/t_{\rm src}<4$, and (III) late decay at $t/t_{\rm src}\geq4$.  The amplitudes and tails are closure dependent, but in both closures the fluctuations decay as the finite source disappears.}
\label{fig:forced-decay}
\end{figure*}

The nonlinear closure used in Figure~\ref{fig:forced-decay} adopts
$D=C_\epsilon K^{3/2}/\ell$, following the turbulent-energy closures of \citet{SchmidtFederrath2011} and \citet{KretschmerTeyssier2020}.  This is a scalar comparison closure, not a complete subgrid-scale model.  With $u'^2=2K$,
\begin{equation}
t_{\rm diss}=\frac{K}{D}
=\frac{\sqrt2\,\ell}{C_\epsilon u'},
\qquad
\frac{t_{\rm diss}}{P_0}
=\frac{\sqrt2}{2\pi C_\epsilon}
\frac{\ell/H}{\mathcal M_{\rm turb}},
\label{eq:tdiss_orbit}
\end{equation}
where $\ell$ is the adopted cascade length, $C_\epsilon$ is the dimensionless dissipation coefficient, $P_0=2\pi/\Omega_0$, and $\mathcal M_{\rm turb}=u'/c_s$.

\begin{figure*}[t]
\centering
\includegraphics[width=0.95\textwidth]{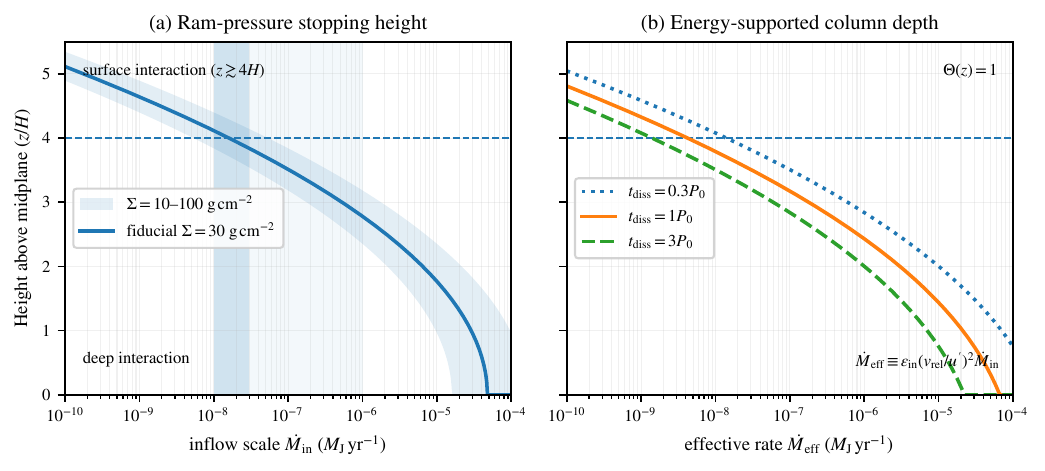}
\caption{Conditional inflow-depth scalings for the inputs in Table~\ref{tab:modelinputs}.  Panel (a) uses Equation~(\ref{eq:zshock}); the bands show planetary accretion rates (HST-derived and model-converted) \citep{Zhou2025}, not measured CPD-boundary inflow.  Panel (b) solves $\Theta(z)=1$ for $t_{\rm diss}=0.3$, 1, and $3P_0$ against the effective coordinate $\dot M_{\rm eff}$.  The bands are not transferred because the mapping from planetary accretion to CPD inflow and the factors $\epsilon_{\rm in}$ and $u'$ remain unsolved.}
\label{fig:infall-depth}
\end{figure*}

For the vertical estimate, the CPD is assigned the isothermal profile
\begin{equation}
\rho_0(z)=\frac{\Sigma}{\sqrt{2\pi}H}
\exp\left(-\frac{z^2}{2H^2}\right),
\qquad
P_{\rm mid}=\frac{\Sigma\Omega_0c_s}{\sqrt{2\pi}},
\label{eq:verticaldisk}
\end{equation}
where $\Sigma$ is the total two-sided gas surface density and $P_{\rm mid}$ is the midplane gas pressure.  For a one-sided footprint $A_{\rm in}$, the fixed-strip areal loading is $\dot\Sigma_{\rm strip}=\dot M_{\rm in}/A_{\rm in}$.  Equating the normal momentum flux $P_{\rm ram}=\dot\Sigma_{\rm strip}|v_{{\rm in},z}|$ to the local gas pressure gives
\begin{equation}
\frac{z_{\rm sh}}{H}
=
\begin{cases}
\left[2\ln\left(P_{\rm mid}/P_{\rm ram}\right)\right]^{1/2},
&0<P_{\rm ram}<P_{\rm mid},\\
0,
&P_{\rm ram}\geq P_{\rm mid}.
\end{cases}
\label{eq:zshock}
\end{equation}
As $P_{\rm ram}\rightarrow0^+$, $z_{\rm sh}/H\rightarrow\infty$.  For $P_{\rm ram}\geq P_{\rm mid}$, setting $z_{\rm sh}=0$ denotes penetration to the deepest layer represented by the one-sided model; it is a clipping convention rather than a pressure-balance solution below the midplane.  Here $v_{{\rm in},z}$ is the component of the incident velocity normal to the disk.  For the illustrative calculation in Figure~\ref{fig:infall-depth}, we set $|v_{{\rm in},z}|=v_{\rm rel}=0.7v_{\rm K}$.  This equality is an adopted scaling, not a general relation between the normal and relative velocities, and Equation~(\ref{eq:zshock}) is only a momentum-balance estimate of the interaction height.

Let $\epsilon_{\rm in}$ be the effective fraction of incident mechanical power converted into disordered motion, including unresolved capture and mechanical coupling, and let $v_{\rm rel}$ be the stream--disk relative speed.  Define
\begin{equation}
\dot M_{\rm eff}
=\epsilon_{\rm in}
\left(\frac{v_{\rm rel}}{u'}\right)^2\dot M_{\rm in}.
\label{eq:mdoteff}
\end{equation}
The one-sided column above $z$ and the ratio of energy supplied in one dissipation time to the fluctuation energy of that column are
\begin{align}
\Sigma_{>}(z)
&=\frac{\Sigma}{2}
\operatorname{erfc}\left(\frac{z}{\sqrt2H}\right),
\label{eq:uppercolumn}\\
\Theta(z)
&=\frac{\dot M_{\rm eff}t_{\rm diss}}
{A_{\rm in}\Sigma_{>}(z)}.
\label{eq:theta}
\end{align}
The deepest energy-supported layer satisfies $\Theta=1$.  With
$\Sigma_{\rm forced}=\dot M_{\rm eff}t_{\rm diss}/A_{\rm in}$, this gives
\begin{equation}
\frac{z_{\rm forced}}{H}
=\begin{cases}
\sqrt2\,\operatorname{erfc}^{-1}
\left(2\Sigma_{\rm forced}/\Sigma\right),
&0<2\Sigma_{\rm forced}/\Sigma<1,\\
0,
&2\Sigma_{\rm forced}/\Sigma\geq1.
\end{cases}
\label{eq:zforced}
\end{equation}
In the zero-forcing limit, $z_{\rm forced}/H\rightarrow\infty$ and no finite layer is supported.  When $\Sigma_{\rm forced}\geq\Sigma/2$, the supplied energy is sufficient for the entire modeled upper half-column, so the solution is clipped at the midplane, $z_{\rm forced}=0$.  Thus weak inflow can perturb a tenuous atmosphere while supplying too little power to maintain fluctuations through the full column.  The quantities $\epsilon_{\rm in}$, $u'$, and the CPD-boundary rate $\dot M_{\rm in}$ are not fixed by a measured planetary accretion rate.

The measured H$\alpha$ rate constrains the planetary sink, not the upstream CPD inflow:
\begin{equation}
\dot M_{\rm in}
=\dot M_p+\dot M_{\rm out,CPD}
+\frac{dM_{\rm CPD}}{dt}.
\label{eq:cpdinfallbudget}
\end{equation}
Here $\dot M_p$ is the rate accreted by the planet, $\dot M_{\rm out,CPD}$ is the rate leaving the CPD control volume, and $dM_{\rm CPD}/dt$ is the rate of change of the stored CPD mass.
Three-dimensional simulations show high-latitude Hill and CPD entry with midplane outflow \citep{TanigawaOhtsukiMachida2012,LiChenLin2023}; late-infall calculations likewise find stronger perturbations above approximately $4H$ than at the midplane \citep{Huhn2026}.  These results motivate the vertical treatment but do not determine $\epsilon_{\rm in}$, $A_{\rm in}$, or the dynamically coupled mass.

\subsection{Reduced stratified calculation}

We use dimensionless height $\zeta=z/H$, time $\tau=\Omega_0t$, and density $\hat\rho=\rho_0H/\Sigma=(2\pi)^{-1/2}e^{-\zeta^2/2}$.  In this subsection, $R_{ij}$ and $K$ are normalized by $c_s^2$.  Retaining the exact Keplerian shear and Coriolis couplings gives
\begin{align}
\frac{\partial R_{xx}}{\partial\tau}
&=4R_{xy}-\Gamma_{\rm cas}R_{xx}+Q+\mathcal D_zR_{xx},
\label{eq:toyRxx}\\
\frac{\partial R_{yy}}{\partial\tau}
&=-2(2-q)R_{xy}-\Gamma_{\rm cas}R_{yy}+Q+\mathcal D_zR_{yy},
\label{eq:toyRyy}\\
\frac{\partial R_{zz}}{\partial\tau}
&=-\Gamma_{\rm cas}R_{zz}+Q+\mathcal D_zR_{zz},
\label{eq:toyRzz}\\
\frac{\partial R_{xy}}{\partial\tau}
&=2R_{yy}-(2-q)R_{xx}-\Gamma_{\rm cas}R_{xy}
+\mathcal D_zR_{xy}.
\label{eq:toyRxy}
\end{align}
The nonlinear cascade and conservative vertical transport are
\begin{align}
\Gamma_{\rm cas}
&=C_\epsilon\frac{\sqrt K}{\ell/H},
\qquad
K=\frac12(R_{xx}+R_{yy}+R_{zz}),
\label{eq:toydamping}\\
\mathcal D_zR
&=\frac{1}{\hat\rho}\frac{\partial}{\partial\zeta}
\left(\hat\rho\widehat\kappa_t
\frac{\partial R}{\partial\zeta}\right).
\label{eq:toytransport}
\end{align}
Here $\Gamma_{\rm cas}$ is the dimensionless cascade rate and $\widehat\kappa_t$ is the dimensionless vertical diffusivity.  The illustrative source deposits equal power into the three diagonal components:
\begin{align}
p_{\rm src}(\tau)
&=\frac{\epsilon_{\rm in}}{2}
\frac{\dot\Sigma_{\rm src}[t(\tau)]}{\Sigma\Omega_0}
\left(\frac{v_{\rm rel}}{c_s}\right)^2,
\label{eq:pin_dimensionless}\\
Q(\zeta,\tau)
&=\frac{2p_{\rm src}(\tau)}{3\hat\rho(\zeta)}
\phi(\zeta;\zeta_{\rm sh}(\tau)),
\label{eq:toyforcing}
\end{align}
where $p_{\rm src}$ is normalized by $\Sigma c_s^2\Omega_0$, $t(\tau)=\tau/\Omega_0$, and $\zeta_{\rm sh}=z_{\rm sh}/H$.  The implemented one-sided deposition kernel is the reflected Gaussian
\begin{equation*}
\phi(\zeta;\zeta_{\rm sh})=
\frac{
\exp[-(\zeta-\zeta_{\rm sh})^2/(2\sigma_\zeta^2)]
+\exp[-(\zeta+\zeta_{\rm sh})^2/(2\sigma_\zeta^2)]
}{\sqrt{2\pi}\sigma_\zeta},
\qquad \sigma_\zeta=0.3.
\end{equation*}
The reflected form has unit integral over $0\leq\zeta<\infty$.  It is not renormalized after truncation at the numerical upper boundary $\zeta_{\max}=5.5$, so forcing centered above the modeled column vanishes from the domain rather than being reassigned to its uppermost layer.  The power actually deposited below the upper boundary is
\begin{equation}
P_{\rm dep}(\tau)
=p_{\rm src}(\tau)\int_0^{\zeta_{\max}}
\phi(\zeta;\zeta_{\rm sh}(\tau))\,d\zeta.
\label{eq:deposited_power}
\end{equation}
Equal diagonal injection is a pedagogical reduction; a three-dimensional calculation must determine the anisotropic stress source.

Both closure integrations begin from the unforced state
\begin{equation*}
R_{xx}(\zeta,0)=R_{yy}(\zeta,0)=R_{zz}(\zeta,0)
=R_{xy}(\zeta,0)=0.
\end{equation*}

The comparison closure replaces only the nonlinear sink with
\begin{equation}
\left.\frac{\partial R_{ij}}{\partial\tau}\right|_{\rm sink}
=-\Gamma_{\rm lin}R_{ij},
\qquad
\Gamma_{\rm lin}=\frac{1}{2\pi}.
\label{eq:linear_decay_closure}
\end{equation}
The linear closure decays exponentially on the timescale $P_0$, whereas the $K^{3/2}$ closure has an algebraic tail.  Table~\ref{tab:modelinputs} gives the closure coefficients, numerical grid, solver tolerances, and resolution-convergence result.

The forced-decay example adopts a prescribed exponentially declining source and one representative change in deposition radius:
\begin{align}
\dot M_{\rm src}(t)
&=\dot M_0e^{-t/t_{\rm src}},\\
\lambda_{\rm ph}(t)
&=\begin{cases}
1/4, & t<t_{\rm src},\\
0.83, & t\geq t_{\rm src},
\end{cases}\\
r_{\rm c}(t)
&=\frac{\lambda_{\rm ph}(t)^2}{3}R_{{\rm H},c},\\
\dot\Sigma_{\rm src}(t)
&=\frac{\dot M_{\rm src}(t)}{\pi r_{\rm c}(t)^2}.
\label{eq:les_gap_history}
\end{align}
Here $\lambda_{\rm ph}(t)$ is the phase-dependent dimensionless inflow angular momentum, $r_{\rm c}(t)$ is its corresponding circularization and deposition radius, and $R_{{\rm H},c}$ is the Hill radius of planet c.  The late value $0.83$ is adopted from the conditional ballistic result in Equation~(\ref{eq:ballistic_result}); it does not determine the source amplitude or transition time.  For this illustration we set $t_{\rm src}=10^4$~yr based on the PDS~70 depletion interval in Equation~(\ref{eq:gap_clearing_comparison}).  The normalization $\dot M_0$ is prescribed, and $\dot M_{\rm src}$ is not a prediction of the three-dimensional Hill or CPD inflow.  Accordingly, $\epsilon_{\rm in}$ remains an effective unresolved factor between the prescribed source power and the mechanically coupled CPD forcing.

The coincidence of the phase switch with $t=t_{\rm src}$ is imposed for illustration; it is not derived from either the depletion history or the ballistic calculation.  For the fiducial values in Table~\ref{tab:modelinputs}, $P_0=4.2$~yr, so the local response can adjust while the prescribed source evolves.

As $\dot M_{\rm src}$ and the corresponding forcing approach zero, the Coriolis-coupled shear exchange does not sustain the fluctuations against the remaining sink.  After orbit averaging and source shutoff at $\tau_s$, the nonlinear envelope in the dimensionless variables of this subsection is
\begin{equation}
K(\tau)=\left[
K(\tau_s)^{-1/2}
+\frac{C_\epsilon}{2(\ell/H)}(\tau-\tau_s)
\right]^{-2}.
\label{eq:post_infall_decay_solution}
\end{equation}
The reduced calculation therefore permits a transient forced phase but provides no sustained late-stage turbulence or permanent global $\alpha$.

\clearpage

\section{Adopted Quantities and Numerical Scalings}\label{app:scalings}
\setcounter{equation}{0}

Table~\ref{tab:inputs} collects the observational and satellite-formation quantities used in the text and figures.  Table~\ref{tab:modelinputs} gives the illustrative model and numerical inputs used for the analytic scalings and original figures.

\begin{table*}[t]
\centering
\caption{Adopted observational and satellite-formation quantities.}
\label{tab:inputs}
\footnotesize
\setlength{\tabcolsep}{5pt}
\renewcommand{\arraystretch}{1.15}
\begin{tabular}{@{}lll@{}}
\toprule
\inputcola{Quantity} & \inputcolb{Adopted value} & \inputcolc{Use or source} \\
\midrule
\inputcola{PDS~70 system} & \inputcolb{$M_*=0.76\,\Msun$, $d=112.4$~pc, age $5.4\pm1.0$~Myr, $a_b=22$~au, $a_c=34.5$~au} & \inputcolc{Architecture, continuum, and chronology; \citet{Muller2018,Keppler2018,Haffert2019,Wang2021}} \\
\inputcola{PDS~70~c mass and geometry} & \inputcolb{$M_c=4$--$12\,\Mjup$, circumstellar-disk inclination $51.7^\circ$, $R_{\rm d}\lesssim1.2$~au} & \inputcolc{Hill scaling and coplanar emitting-area conversion; \citet{Keppler2019,Benisty2021,Shibaike2026}} \\
\inputcola{Discovery continuum inputs} & \inputcolb{$F_{855}=86\pm16\,\mu{\rm Jy}$, $T_{\rm d}=26$~K, $\kappa_{855}=0.79$ and $3.63~{\rm cm^2~g^{-1}}$} & \inputcolc{Thin-mass and conditional isothermal thick-area examples; \citet{Benisty2021,Birnstiel2018}} \\
\inputcola{Follow-up continuum constraints} & \inputcolb{$\alpha_{\rm B4,B7}=2.01\pm0.22$, $M_{\rm d}=0.008$--$0.063\,\Mearth$, ring background temperature 22~K} & \inputcolc{\citet{DominguezJamett2025,Fasano2025,Shibaike2026}} \\
\inputcola{0.88-mm disk--host relation} & \inputcolb{slope 1.54, intercept 1.88, intrinsic scatter 0.74~dex} & \inputcolc{Equation~(\ref{eq:wu_relation}); \citet{Wu2020}} \\
\inputcola{SR~12~c continuum inputs} & \inputcolb{$M_p=16\pm2\,\Mjup$, $d=139\pm5$~pc, $F_{0.88}=127\pm14\,\mu{\rm Jy}$} & \inputcolc{Figure~\ref{fig:host-scaling}; \citet{Wu2022,Finley2026}} \\
\inputcola{SR~12~c disk and accretion} & \inputcolb{$\dot M_p=(8\pm2)\times10^{-12}\,\Msun\,{\rm yr^{-1}}$, $M_{\rm d}=0.007$--$0.03\,\Mearth$, $M_{\rm g,acc}\sim0.03\,\Mjup$} & \inputcolc{Accretion measurement and gas inventory; the dust masses are the published \citet{Wu2022} model values at their adopted distance; \citet{Finley2026}} \\
\inputcola{Callisto formation timescale} & \inputcolb{$\sim10^6$~yr} & \inputcolc{\SEMM{} gas-drag clearing calculation; \citet{MosqueiraEstrada2003a,Estrada2009,Mosqueira2010SSR}} \\
\inputcola{Iapetus formation timescale} & \inputcolb{$\sim10^7$~yr} & \inputcolc{\SEMM{} gas-drag clearing calculation; \citet{MosqueiraEstrada2003b,Estrada2009,Mosqueira2010SSR}} \\
\bottomrule
\end{tabular}
\end{table*}

\begin{table*}[t]
\centering
\caption{Illustrative common-gap, ballistic, satellitesimal, and inflow-forcing inputs.}
\label{tab:modelinputs}
\footnotesize
\setlength{\tabcolsep}{5pt}
\renewcommand{\arraystretch}{1.15}
\begin{tabular}{@{}lll@{}}
\toprule
\inputcola{Quantity} & \inputcolb{Adopted value} & \inputcolc{Use or source} \\
\midrule
\inputcola{Planet architectures} & \inputcolb{Jupiter--Saturn: $M_*=1\,\Msun$, $(M_b,a_b)=(1.00\,\Mjup,5.20~{\rm au})$, $(M_c,a_c)=(0.299\,\Mjup,9.58~{\rm au})$; PDS~70-like: $M_*=0.76\,\Msun$, $(M_b,a_b)=(5\,\Mjup,22~{\rm au})$, $(M_c,a_c)=(7.5\,\Mjup,34.5~{\rm au})$} & \inputcolc{Common-gap comparison; nominal Solar System values and \citet{Muller2018,Keppler2018,Haffert2019,Trevascus2025,Shibaike2026}} \\
\inputcola{Gap closure and reservoir} & \inputcolb{$h_{\rm J}=0.0495$, $h_{\rm S}=0.0577$, $h_b=0.0725$, $h_c=0.0811$, $C_T=0.01$, $C_{H,b}=C_{H,c}=0.1$, $w_p=2.5R_{{\rm H},p}$} & \inputcolc{Direct-irradiation profiles and illustrative closure; the width follows \citet{MassetDAngeloKley2006}, while $C_H$ requires three-dimensional calibration} \\
\inputcola{Ballistic ensemble} & \inputcolb{$L_1$ only, $h_c=0.08$ (rounded), 17 heights over $-H_c\leq z\leq H_c$, $0.1\leq u_n\leq0.5$, $|u_t|\leq0.1$, $u_z=0$, $S_{\rm sh}=0.5\RH$} & \inputcolc{Gaussian vertical-density and normal-flux weighting; rotating-frame velocities in units of $\Omega_p\RH$; normalized profile in Section~\ref{sec:ballistic}} \\
\inputcola{Controlled satellite-embryo drift} & \inputcolb{For Jupiter, Saturn, PDS~70~b, and PDS~70~c: one Callisto mass of solids uniformly inside $R_{{\rm H},p}/3$, gas-to-solids ratio 100, evaluation at $r=R_{{\rm H},p}/10$, $b_{\rm iso}=5$, $\rho_s=1~{\rm g~cm^{-3}}$, $\mu=2.34$, $d\ln P/d\ln r=-3/2$, $C_D=0.44$, $C_I=1/2.7$} & \inputcolc{Isolation, gas-drag, and Type-I displacement estimates in Equations~(\ref{eq:controlled_satellite_columns})--(\ref{eq:isolation_torque_time}) and Table~\ref{tab:controlled-embryo-drift}; temperatures from Figure~\ref{fig:passivetemperature}; \citet{Tanaka2002}} \\
\inputcola{Local CPD scaling} & \inputcolb{$M_p=7.5\,\Mjup$, $R_0=0.5$~au, $P_0=4.2$~yr, $H/R_0=0.1$, $\Sigma=30~{\rm g~cm^{-2}}$, $|v_{{\rm in},z}|=v_{\rm rel}=0.7v_{\rm K}$} & \inputcolc{Figures~\ref{fig:infall-depth} and \ref{fig:forced-decay}; the depth calculation also spans $\Sigma=10$--$100~{\rm g~cm^{-2}}$, and the velocity equality is an illustrative scaling} \\
\inputcola{Forcing geometry and times} & \inputcolb{$A_{\rm in}=2\pi R_0H$, $A_{\rm dep}(t)=\pi r_{\rm c}(t)^2$, $t_{\rm diss}=0.3$, 1, and $3P_0$, $\dot M_0=5.5\times10^{-5}\,\Mjup\,{\rm yr}^{-1}$, $t_{\rm src}=10^4$~yr, $\dot M_{\rm eff}=\epsilon_{\rm in}(v_{\rm rel}/u')^2\dot M_{\rm in}$} & \inputcolc{Prescribed illustrative history for Appendix~\ref{app:les}; $t_{\rm src}$ is based on the PDS~70 depletion interval in Equation~(\ref{eq:gap_clearing_comparison}); Figure~\ref{fig:infall-depth} uses fixed-strip loading and Figure~\ref{fig:forced-decay} uses the prescribed deposition-area loading} \\
\inputcola{Reduced stress closure} & \inputcolb{$q=3/2$, $C_\epsilon=1$, $\ell=0.5H$, $\widehat\kappa_t=0.01$, deposition width $0.3H$, $\epsilon_{\rm in}=10^{-3}$} & \inputcolc{The scalar $K^{3/2}$ sink follows \citet{SchmidtFederrath2011,KretschmerTeyssier2020}; other coefficients are illustrative} \\
\inputcola{Numerical integration} & \inputcolb{$0\leq\zeta\leq5.5$, 111 nodes, DOP853, relative and absolute tolerances $10^{-8}$ and $10^{-12}$, $\Delta\tau_{\max}=1$} & \inputcolc{Zero-flux boundaries; the 56-to-111-node convergence test changes sampled diagnostics by at most 1.7\%} \\
\inputcola{Planetary accretion comparison} & \inputcolb{$(1\text{--}3)\times10^{-8}\,\Mjup\,{\rm yr^{-1}}$ for the 2024 HST measurement; $10^{-8}$--$10^{-6}\,\Mjup\,{\rm yr^{-1}}$ across published conversions} & \inputcolc{Shown only as an empirical comparison, not identified with CPD inflow; \citet{Zhou2025}} \\
\bottomrule
\end{tabular}
\end{table*}

\subsection{Convenient continuum forms}

At 855-$\mu$m, Equation~(\ref{eq:thinmass}) can be written
\begin{multline}
M_{\rm d}=0.031\,\Mearth
\left(\frac{F_{855}}{86\,\mu{\rm Jy}}\right)
\left(\frac{d}{112.4~{\rm pc}}\right)^2\\
\times\left(\frac{0.79~{\rm cm^2~g^{-1}}}{\kappa_{855}}\right)
\frac{B_{855}(26~{\rm K})}{B_{855}(T_{\rm d})}.
\label{eq:massscaled}
\end{multline}
For an optically thick source, the equivalent circular projected radius is
\begin{multline}
R_{\rm proj}=0.46~{\rm au}
\left(\frac{f_{\rm d}F_{855}}{86\,\mu{\rm Jy}}\right)^{1/2}
\left(\frac{d}{112.4~{\rm pc}}\right)\\
\times\left[\frac{B_{855}(26~{\rm K})}{B_{855}(T)}\right]^{1/2}.
\label{eq:rprojscaled}
\end{multline}
For a circular disk inclined by $i$, $R_{\rm phys}=R_{\rm proj}/\sqrt{\cos i}$.  At $i=51.7^\circ$, 26~K, and $f_{\rm d}=1$, $R_{\rm phys}=0.58$~au.  This scaling is restricted to the uniform isothermal case; Equation~(\ref{eq:thickradial}) applies to a radial temperature profile.

\subsection{Hill and circularization-radius scalings}

For a convenient $8\,\Mjup$ normalization near the $7.5\,\Mjup$ fiducial used in the figures, PDS~70~c has
\begin{equation}
\RH=5.2~{\rm au}
\left(\frac{a_p}{34.5~{\rm au}}\right)
\left(\frac{M_p}{8\,\Mjup}\right)^{1/3}
\left(\frac{0.76\,\Msun}{M_*}\right)^{1/3}.
\label{eq:hillscaled}
\end{equation}
The pre-gap and gap-fed circularization scales are therefore
\begin{align}
r_{\rm c,pre}&=0.11~{\rm au}
\left(\frac{\RH}{5.2~{\rm au}}\right),\\
r_{\rm c,gap}&=1.7~{\rm au}
\left(\frac{\RH}{5.2~{\rm au}}\right).
\label{eq:rcscaled}
\end{align}
The wide architecture therefore permits an au-scale CPD.  Its longer orbital period does not by itself imply slower gap clearing.  In the adopted closure, the larger PDS~70 planet-to-star mass ratios strengthen the tidal torque, while the larger aspect ratios weaken it through the $h_p^{-3}$ dependence.  These effects substantially offset one another: Equations~(\ref{eq:gap_clearing_js})--(\ref{eq:gap_clearing_comparison}) give clearing and depletion timescales of order $10^4$~yr for both architectures, and Table~\ref{tab:thermal-clearing} shows that the comparison persists across the representative matched temperature profiles.

\section{Construction of the Retained Figures and Numerical Checks}\label{app:figures}

Figures~\ref{fig:architecture}--\ref{fig:forced-decay} were constructed for this paper.  Figure~\ref{fig:architecture} shows the two architectures in Table~\ref{tab:modelinputs} and the Hill radii calculated from Equation~(\ref{eq:hill}); its common-gap shading is schematic rather than a fitted gap boundary.  Figure~\ref{fig:sed} shows the flux compilation in Table~\ref{tab:sed}.  Figure~\ref{fig:massT} shows the evaluation of Equation~(\ref{eq:thinmass}).  Figure~\ref{fig:passivetemperature} shows the direct-irradiation relation in Equation~(\ref{eq:directtemperature}) and the matched passive two-layer constructions described in Section~\ref{sec:passivetemperature}; the literature comparison points and the adopted \SEMM{} temperatures are identified in its caption.  Figure~\ref{fig:cpdtemperaturemodels} shows the passive star- and planet-heated circumplanetary profiles described in Section~\ref{sec:cpdoutertemperature}; its literature temperatures and the $\RH/48$ and $\RH/3$ reference scales are likewise identified in its caption.  Figure~\ref{fig:host-scaling} shows the values calculated from Equations~(\ref{eq:flux140}) and (\ref{eq:wu_relation}) using the Taurus, Chamaeleon~I, and Lupus comparison samples digitized from the published vector figure of \citet{Wu2020}, the six companion limits tabulated there, the WISPIT~2~b limit from \citet{Facchini2026}, and the measured PDS~70~c and SR~12~c fluxes.  Figure~\ref{fig:hill} shows the values calculated from Equations~(\ref{eq:hill}), (\ref{eq:rcpre}), and (\ref{eq:rcgap}).  Figure~\ref{fig:gap-cpd-loading} shows the mapping of the trajectory result in Equation~(\ref{eq:ballistic_result}) through Equation~(\ref{eq:ballistic_surface_density}).  Figure~\ref{fig:infall-depth} shows the evaluation of Equations~(\ref{eq:zshock}) and (\ref{eq:zforced}), and Figure~\ref{fig:forced-decay} shows the integration of Equations~(\ref{eq:toyRxx})--(\ref{eq:toyRxy}) with the prescribed source and deposition history in Equation~(\ref{eq:les_gap_history}).

\subsection{Ballistic profile checks}

The ballistic profile retained in this paper is defined by the illustrative ensemble specified in Section~\ref{sec:ballistic}.  Distances are normalized by $\RH$, velocities by $\Omega_p\RH$, and time by $\Omega_p^{-1}$.  The $+x$ direction points from the planet toward the star, $+z$ follows the orbital angular momentum, and $+y$ completes the right-handed rotating frame.  With $s=(x^2+y^2+z^2)^{1/2}$ and dots denoting derivatives with respect to the dimensionless time $\Omega_pt$, the integrated three-dimensional Hill equations are
\begin{equation}
\begin{aligned}
\ddot x&=2\dot y+3x-\frac{3x}{s^3},\\
\ddot y&=-2\dot x-\frac{3y}{s^3},\\
\ddot z&=-z-\frac{3z}{s^3}.
\end{aligned}
\label{eq:ballistic_hill_equations}
\end{equation}
The vertical launch column at $x=1$ represents the starward $L_1$ entry region.  For each sampled height $z_i$, the exact initial state is
\begin{equation}
(x,y,z,\dot x,\dot y,\dot z)_0
=(1,0,z_i,-u_n,u_t,0),
\label{eq:ballistic_initial_state}
\end{equation}
where $u_n>0$ is the inward-normal speed and $u_t$ is positive along $+y$.  The adopted ensemble contains only $L_1$ launches.  The corresponding hypothetical $L_2$ ensemble is defined by the symmetry
\begin{equation*}
(x,y,z,\dot x,\dot y,\dot z)
\longrightarrow(-x,-y,z,-\dot x,-\dot y,\dot z),
\end{equation*}
which maps Equation~(\ref{eq:ballistic_initial_state}) to $(-1,0,z_i,u_n,-u_t,0)$ and preserves both the Hill equations and the dimensionless axial angular momentum.  Thus an independently normalized $L_2$ ensemble with the reflected entry distribution would have the same circularization distribution; it is not included in the adopted interplanetary-reservoir calculation.

At the starward $L_1$ entry region, a $17\times17\times17$ grid samples 17 heights over $-H_c\leq z\leq H_c$, $0.1\leq u_n\leq0.5$, and $|u_t|\leq0.1$, with $u_z=0$.  The 4913 launched trajectories are integrated with a fourth-order Runge--Kutta method at a dimensionless step of $10^{-3}$.  An integration ends at the first crossing of $S_{\rm sh}=0.5\RH$, after 8000 steps ($8\Omega_p^{-1}$), or when the trajectory reaches $1.2\RH$ after the first 50 steps; 4205 trajectories reach the interaction surface.  Their quadrature weights are proportional to $u_n\exp[-z^2/(2H_c^2)]$, with half weight at the two endpoints of the vertical grid.  The corresponding dimensionless inertial angular momentum is $\lambda=xv_y-yv_x+x^2+y^2$.  Equation~(\ref{eq:ballistic_mapping}) maps the resulting $j_z$ distribution to $x=R_{\rm c}/\RH$, and Equation~(\ref{eq:ballistic_surface_density}) maps its normalized distribution $p_x(x)$ into $\widehat{\Sigma}(x)$ without subsequent redistribution.

To construct the plotted local profile, we form a 600-bin mass-flux-weighted histogram over $x_{\min}=\max[0.05,\min(x_i)-0.03]$ to $x_{\max}=\min[0.36,\max(x_i)+0.03]$, where $x_i=R_{{\rm c},i}/\RH$.  The histogram density is convolved with a Gaussian of standard deviation six bins, using nearest-value extension at the two boundaries, and normalized by trapezoidal integration.  Because this smoothing slightly changes the nonlinear moment $\langle\sqrt{3x}\rangle$, the positive smoothed density $p_0(x)$ is given the exponential tilt
\begin{equation*}
\begin{aligned}
p_\beta(x)&=\frac{p_0(x)}{Z(\beta)}
\exp\{\beta[\sqrt{3x}-\langle\lambda\rangle]\},\\
Z(\beta)&=\int p_0(x')
\exp\{\beta[\sqrt{3x'}-\langle\lambda\rangle]\}\,dx'.
\end{aligned}
\end{equation*}
The coefficient $\beta$ is determined by bisection so that $\int\sqrt{3x}\,p_\beta(x)\,dx=\langle\lambda\rangle$; the adopted finite-height profile has $\beta=0.00170$.  The plotted curve is $\widehat{\Sigma}=p_\beta/x$ and is displayed where $\widehat{\Sigma}>0.03$.  Direct integration recovers both $\int p_\beta\,dx=1$ and the independently calculated unrounded flux-weighted moment $\langle\lambda\rangle=0.8291667$ to machine precision.  The dashed comparison in Figure~\ref{fig:gap-cpd-loading} is reconstructed from the earlier midplane ensemble using the same 600-bin, six-bin-Gaussian-smoothed, moment-matching prescription.

Halving or doubling the integration step leaves all 4205 successful crossings unchanged and changes $\langle R_{\rm c}/\RH\rangle$ by less than 0.04\%.  Moving the illustrative interaction surface from $0.4\RH$ to $0.6\RH$ gives mean radii of $0.222\RH$ and $0.245\RH$, respectively.  Narrowing the sampled velocities to $0.15\leq u_n\leq0.45$ and $|u_t|\leq0.05$, or widening them to $0.05\leq u_n\leq0.55$ and $|u_t|\leq0.15$, gives $0.228\RH$ and $0.232\RH$.  Doubling the maximum integration duration to $16\Omega_p^{-1}$ leaves the 4205 crossings and mean radius unchanged.  Repeating that longer integration with escape boundaries at $1.5\RH$, $2\RH$, and $5\RH$ also leaves the result unchanged: in every case the other 708 trajectories cross the escape boundary and none remain active at the time limit.  For the Jacobi integral
\begin{equation*}
C_J=3x^2-z^2+\frac{6}{s}
-(\dot x^2+\dot y^2+\dot z^2),
\end{equation*}
the baseline successful trajectories have a maximum absolute drift of $9.1\times10^{-11}$ and a maximum relative drift of $1.2\times10^{-11}$.  These checks establish the numerical stability and representative radial scale of the specified ensemble; they do not turn its velocity prior, interaction surface, or central $\pm H_c$ column into a prediction of the three-dimensional flow.  The calculation does not determine when $L_1$ dominates the supply, nor the supplied mass or the mass retained.

\subsection{Reduced stress calculation checks}

The stress calculation uses the grid and tolerances in Table~\ref{tab:modelinputs}.  At every sampled time, the normal stresses remain nonnegative and the horizontal covariance satisfies $R_{xy}^2\leq R_{xx}R_{yy}$.  Repeating the calculation with 56 rather than 111 vertical nodes changes the column, surface, and midplane diagnostics at $t/t_{\rm src}=1$, 4, 8, and 12 by at most 1.7\%.  This verifies the numerical resolution required for the regime comparison in Figure~\ref{fig:forced-decay}; it does not remove the physical dependence on the illustrative closure or unresolved inflow coupling.

\section{Methodological Note on Inherited Bias in Artificial-intelligence-assisted Scientific Reasoning}\label{app:ai-bias}

This manuscript was prepared with editorial assistance from artificial intelligence (AI), and the process exposed a methodological failure worth recording.  When asked to summarize the observational evidence for disk turbulence, the AI assistant repeatedly reconstructed a conventional literature framing in which turbulence was treated as the default state and the observations were asked to rule it out.  It did so even after the intended evidentiary burden had been stated explicitly.  This was an inherited interpretive bias likely produced by the repeated association, in the scientific literature used to train language models, of nonthermal line widths with turbulence, turbulence with angular-momentum transport, and angular-momentum transport with an $\alpha$-disk closure.

\bibliography{references}
\bibliographystyle{aasjournal}

\end{document}